\documentclass[a4paper,11pt]{article}

\usepackage{tabstackengine,threeparttable}
\usepackage[inline]{enumitem}
\RequirePackage{amsthm,amsmath,amsfonts,amssymb,mathtools,titling,authblk,setspace}
\RequirePackage[authoryear]{natbib}

\usepackage[margin=1.34in]{geometry}

\RequirePackage[colorlinks,citecolor=blue,linkcolor=blue,urlcolor=blue]{hyperref}
\mathtoolsset{mathic=true} 

\numberwithin{equation}{section}

\theoremstyle{plain}
\newtheorem{theorem}{Theorem}
\newtheorem{lemma}{Lemma}
\newtheorem{proposition}{Proposition}

\theoremstyle{remark}
\newtheorem{remark}{Remark}
\newtheorem{definition}{Definition}

\newcommand{\mynegspace}{\hspace{-0.12em}}
\newcommand{\lvvvert}{\rvert\mynegspace\rvert\mynegspace\rvert}
\newcommand{\rvvvert}{\rvert\mynegspace\rvert\mynegspace\rvert}
\newcommand{\dotrightarrow}{\mathrel{\mathrlap{\bullet}{\rightarrow}}}
\newcommand{\bigzero}{\scalebox{0.9}{\mbox{\large $0$}}}

\DeclareMathOperator{\col}{col}
\newcounter{counter:items}

\begin{document}
\emergencystretch 3em

\title{The general solution to an autoregressive law of motion}

\author[1]{Brendan K.\ Beare}
\author[2]{Massimo Franchi}
\author[3]{Phil Howlett}
\affil[1]{School of Economics, University of Sydney}
\affil[2]{Department of Statistical Science, Sapienza University of Rome}
\affil[3]{Centre for Industrial and Applied Mathematics, University of South Australia}

\maketitle

\begin{center}
Accepted for publication in \textit{Quantitative Economics}.
\end{center}
\medskip

\begin{abstract}
We provide a complete description of the set of all solutions to a vector autoregressive law of motion. Every solution is shown to be the sum of three components, each corresponding to a directed flow of time. One component flows forward from the arbitrarily distant past; one flows backward from the arbitrarily distant future; and one flows outward from time zero. The three components are obtained by applying three complementary spectral projections to the solution, these corresponding to a separation of the eigenvalues of the autoregressive coefficient matrix according to whether they are inside, outside or on the unit circle. We establish a one-to-one correspondence between the set of all solutions and a finite-dimensional space of initial conditions.
\end{abstract}

\onehalfspacing

\section{Introduction}\label{sec:intro}

This article describes a general solution procedure for a vector autoregressive law motion. A vector autoregressive law of motion---referred to more simply as an autoregressive law of motion in what follows---is an infinite system of linear equations determined by two objects: a real $N\times N$ matrix $\Phi$ called the autoregressive coefficient and a sequence $\varepsilon=(\varepsilon_t)$ in $\mathbb R^N$ called the innovation sequence and indexed by $t\in\mathbb Z$. The corresponding infinite system of linear equations is
\begin{equation}
	x_t=\Phi x_{t-1}+\varepsilon_t,\quad t\in\mathbb Z.\label{eq:VAR1}
\end{equation}
When a sequence $x=(x_t)$ in $\mathbb R^N$, also indexed by $t\in\mathbb Z$, satisfies \eqref{eq:VAR1} we say that $x$ is a solution to \eqref{eq:VAR1}, or simply a solution. In this article we provide a complete characterization of the set of all solutions. Our particular contribution is to show that every solution can be expressed as a sum of separate self-contained forward, backward and outward flows. We place no restrictions on $\Phi$ and require only that $\varepsilon$ satisfies
\begin{equation}
	\sum_{t\in\mathbb Z}r^{\lvert t\rvert}\lVert\varepsilon_t\rVert<\infty\quad\text{for each }r\in(0,1),\label{eq:subexponential}
\end{equation}
where $\lVert\cdot\rVert$ is the Euclidean norm on $\mathbb R^N$. We will call any sequence in $\mathbb R^N$ that satisfies the condition placed on $\varepsilon$ in \eqref{eq:subexponential} a subexponential sequence. The sequence notation $y=(y_t)$ will always be used to refer to a two-sided sequence $y$ indexed by $t\in\mathbb Z$. 

A few words on terminology and generality are in order. In discussions of the autoregressive law motion the term \emph{innovation sequence} generally refers to a \emph{random} sequence in $\mathbb R^N$, often assumed to be white noise. Such a random sequence $\varepsilon$ is defined on an underlying probability space, say $\Omega$. By fixing a point $\omega\in\Omega$ one obtains a \emph{nonrandom} sequence $\varepsilon(\omega)=(\varepsilon_{t}(\omega))$ in $\mathbb R^N$ which may be called a \emph{realized} innovation sequence. The autoregressive law of motion \eqref{eq:VAR1} is assumed to hold for all realized innovation sequences, i.e.\ for each $\omega\in\Omega$, and thus the set of solutions must also be understood to correspond to a specific choice of $\omega\in\Omega$. The default perspective adopted in this article is that the choice of $\omega\in\Omega$ has already been fixed, so that we are working with a realized innovation sequence $\varepsilon(\omega)$, which we refer to more simply as an innovation sequence and denote by $\varepsilon$. Probabilistic concepts are formally relevant to our solution procedure only insofar as they can be used to justify the subexponential condition imposed on realized innovations. This condition holds very generally. We show in Appendix \ref{sec:prob1} that a sufficient condition for a random sequence $\varepsilon$ in $\mathbb R^N$ to be subexponential for almost every $\omega\in\Omega$ is that the corresponding sequence of expected norms $(\mathrm{E}\lVert\varepsilon_{t}\rVert)$ is subexponential. Thus it suffices to exclude cases where the expected norm of innovations grows exponentially as time progresses or regresses. In particular, any innovation sequence with a time-invariant covariance matrix, and thus any white noise (in either the weak or strong sense), is almost surely subexponential. Almost sure subexponentiality is a two-sided version of the defining property of a subexponential process as stated in \citet[pp.~636--7]{AlSadoon2018}.

The law of motion \eqref{eq:VAR1} is commonly called a VAR(1) law of motion, and when $\varepsilon$ is the realization of white noise a solution to \eqref{eq:VAR1} is commonly called a VAR(1) process. Such processes fall naturally within the scope of our analysis. Our analysis extends easily to higher-order autoregressive laws of motion, i.e.\ VAR($p$) laws of motion, by rewriting them as VAR(1) laws of motion in a space of higher dimension via, for instance, the companion form. Constant or trending deterministic components may be included as additional terms on the right-hand side of an autoregressive law of motion within our framework by regarding them as part of the innovation sequence, provided that trends do not grow exponentially as time progresses or regresses. Moving averages of white noise are almost surely subexponential, so VARMA($p,q$) processes also fall within the scope of our analysis. The autoregressive coefficient is unrestricted, and in particular may have one or more eigenvalues on the unit circle, so integrated processes in the VARIMA($p,d,q$) class are also encompassed by our framework.

Our separation of every solution into a sum of separate self-contained forward, backward and outward flows is accomplished using three real $N\times N$ projection matrices called spectral projections. The three spectral projections sum to the $N\times N$ identity matrix $I$ and are constructed from a basis of generalized eigenvectors of $\Phi$ as described in Section \ref{sec:spectralprojection}. The spectral projections are not in general orthogonal projections. By applying each of the spectral projections to a solution $x$ we obtain three sequences that sum to $x$. We refer to these three sequences as the forward, backward and outward components of $x$. They correspond to the eigenvalues of $\Phi$ that are respectively inside, outside and on the unit circle.

The forward, backward and outward components of $x$ may each be further separated into a sum of two parts. One part is uniquely determined by the behavior of $x$ at the origin of the relevant flow of time (i.e.\ by the \emph{initial} behavior of $x$), and the other part is uniquely determined by $\varepsilon$ and is the same for every solution. We call the former part a predetermined forward, backward or outward $x$-flow and call the latter part a forward, backward or outward $\varepsilon$-flow. Thus every solution $x$ takes the general form
\begin{align}\label{eq:figure}
	x&=\begin{dcases}\text{forward component}&=\begin{dcases}\text{predetermined forward } x\text{-flow}\\[-1.5ex]\quad\quad+\\[-1.5ex]\text{forward }\varepsilon\text{-flow}\end{dcases}\\[-1.5ex]\quad\quad+&\\[-1.5ex]\text{backward component}&=\begin{dcases}\text{predetermined backward }x\text{-flow}\\[-1.5ex]\quad\quad+\\[-1.5ex]\text{backward }\varepsilon\text{-flow}\end{dcases}\\[-1.5ex]\quad\quad+&\\[-1.5ex]\text{outward component}&=\begin{dcases}\text{predetermined outward }x\text{-flow}\\[-1.5ex]\quad\quad+\\[-1.5ex]\text{outward }\varepsilon\text{-flow.}\end{dcases}\end{dcases}
\end{align}
We postpone a precise explanation of the directed flow terminology and of the concept of predetermination to Section \ref{sec:arrow}. For now we offer the following intuitive description of the six flows in \eqref{eq:figure}.
\begin{enumerate}[label=\upshape(\roman*)]
	\item The predetermined forward $x$-flow is a sequence determined by $x$-values in the arbitrarily distant past. This flow converges exponentially to zero as time progresses.
	\item The forward $\varepsilon$-flow is a sequence whose current value is a weighted average of all current and past $\varepsilon$-values. The contribution of each individual $\varepsilon$-value diminishes exponentially as time progresses.
	\item The predetermined backward $x$-flow is a sequence determined by $x$-values in the arbitrarily distant future. This flow converges exponentially to zero as time regresses.
	\item The backward $\varepsilon$-flow is a weighted average of all future $\varepsilon$-values. The individual contributions diminish exponentially as time regresses.
	\item The predetermined outward $x$-flow is a sequence determined by the $x$-value at time zero.
	\item The outward $\varepsilon$-flow is a sequence whose current value is a weighted average of the $\varepsilon$-values between time zero and the current time.
\end{enumerate}
Explicit formul{\ae} for all six flows are provided in Section \ref{sec:mainresult}. In the univariate case, treated in Section \ref{sec:univariate}, these formul{\ae} reduce to well-known expressions not involving spectral projection. The separation of a multivariate solution to \eqref{eq:VAR1} into its forward, backward and outward components via spectral projection, and the further separation of each of these components into a term depending only on the innovations and a term depending only on the choice of an initial condition, is the core contribution of this article.

The flow decomposition in \eqref{eq:figure} is consistent with the discussion in the opening pages of \citet[pp.~9--12]{HannanDeistler1988}. It is explained there that, in cases where $\Phi$ has no eigenvalues on the unit circle and $\varepsilon$ is a stationary process with finite expected norm, there is a unique stationary solution $\tilde{x}$ and the set of all solutions is given by $\tilde{x}+y$ where $y$ is any solution to the corresponding homogeneous difference equation with $\varepsilon=(0)$. The set of all such sequences $y$ is, in general, infinite except in cases where $\Phi$ is nilpotent. We elaborate on the connection between our flow decomposition and the discussion in \citet{HannanDeistler1988} in Remark \ref{rem:HD}.

Misleading or incomplete statements about the set of all solutions can be found in some well-known econometrics textbooks. There are two distinct sources of confusion, one relating to the stationarity of solutions when $\Phi$ has all eigenvalues inside the unit circle, and the other relating to the stationarity of solutions when $\Phi$ has one or more eigenvalues outside the unit circle. We elaborate further in Remarks \ref{rem:BHansen} and \ref{rem:Potscher}.

The outward component of an autoregressive process---i.e., of a solution to \eqref{eq:VAR1} when $\varepsilon$ is a white noise process---is the central concern of the voluminous econometric literature on unit roots and co-integration, though its outward character has rarely been recognized. The central result on the structure of the outward component is known as the Granger or Granger-Johansen representation theorem. A version of this result first appeared in \cite{Granger1986} and \cite{EngleGranger1987}, but contained a flaw related to the possibility of the generalized eigenspace associated with a unit eigenvalue of the autoregressive operator not admitting a basis of eigenvectors, leading to much confusion in subsequent literature. Closely related research reported in \citet{Johansen1988,Johansen1991,Johansen1992,Johansen1995} avoided this problem, but the issue was not explicitly pointed out until \citet{Johansen2008}, where a counterexample to Lemma 1 in \citet{EngleGranger1987} was provided in a footnote. See also \cite{Howlett1982}, which addressed a similar issue in an input retrieval problem, and the related comment by Johansen in \citet[p.~8]{MosconiParuolo2022}. Research on the outward component of an autoregressive process initially focused on cases where there is a unit eigenvalue and all other eigenvalues are inside the unit circle, but beginning with \citet{EngleGrangerHallman1989} eigenvalues anywhere on the unit circle were permitted, these being described as seasonal unit roots when forming conjugate pairs. Further contributions to the study of seasonal unit roots include \citet{HyllebergEngleGrangerYoo1990}, \citet{Gregoir1999a,Gregoir1999b}, \citet{JohansenSchaumburg1999} and, more recently, \citet{BauerWagner2012}.

With a handful of exceptions, nearly all published research on the structure of the outward component of an autoregressive process has obscured its outward character by indexing time with the nonnegative integers. The first exceptions may be \citet{GregoirLaroque1993,GregoirLaroque1994}, where the outward flow of time is not explicitly commented upon but can be recognized through the application of a two-sided cumulation operator to innovations indexed by all integer times. Other articles following this approach include \citet{Gregoir1999a,Gregoir1999b}, \citet{BauerWagner2012} and \citet{FranchiParuolo2019,FranchiParuolo2020,FranchiParuolo2021}. The apparent reluctance of econometricians to allow the outward component of an autoregressive process to be indexed by all integer times may stem from a mistrust of non-causal processes, as the outward $\varepsilon$-flow in \eqref{eq:figure} must necessarily depend on future innovations at negative times. Non-causal autoregressive processes have nevertheless received considerable attention in recent econometric literature, particularly in applications involving rational expectations or speculative price bubbles. See, for instance, \cite{LanneSaikkonen2013,HencicGourieroux2015,GourierouxJasiak2016,GourierouxZakoian2017,AlSadoon2018} and \cite{DavisSong2020}. In particular, \cite{GourierouxZakoian2017} show that a stationary anti-causal autoregressive process may sometimes be given a causal interpretation, and may exhibit locally explosive behavior despite being stationary. Irrespective of the recent interest in empirical applications of non-causal models, the general representation of an autoregressive process as the sum of its forward, backward and outward components reveals the fundamental structure of this class of processes and a pleasing three-way symmetry between three arrows of time.

The statement and proof of the main result of this article, Theorem \ref{theorem:multivariate} in Section \ref{sec:mainresult}, require two tools from linear algebra that may be unfamiliar to many econometricians: spectral projection and the Drazin inverse. Sections \ref{sec:spectralprojection} and \ref{sec:Drazin} respectively introduce spectral projection and the Drazin inverse and summarize properties used in the proof of Theorem \ref{theorem:multivariate}. The essential background required for a sound understanding of both concepts is familiarity with the Jordan form of a square matrix and the associated generalized eigenspace decomposition of $\mathbb C^N$; see e.g.\ \citet[ch.~8]{Axler2024}. For this reason, Theorem \ref{theorem:multivariate} may be accessible to students of statistics or econometrics with a strong background in undergraduate linear algebra.

The remainder of our article is structured as follows. We commence in Section \ref{sec:univariate} with a discussion of the univariate case. Propositions \ref{theorem:AR1forward}--\ref{theorem:AR1outward} respectively concern the cases where the scalar autoregressive coefficient is less then, greater than, or equal to one in magnitude. These results are not substantively novel but serve to introduce the six flows in \eqref{eq:figure} in the simplest possible setting and build intuition in advance of the statement and proof of Theorem \ref{theorem:multivariate} in Section \ref{sec:multivariate}. Section \ref{sec:remarks} contains a series of remarks on Theorem \ref{theorem:multivariate} and its relation to past literature. In Section \ref{sec:arrow} we use the concept of measurability to elaborate upon the terminology used for the six flows in \eqref{eq:figure}. Appendix \ref{sec:prob1} establishes a weak sufficient condition for a random innovation sequence to be subexponential with probability one, thereby facilitating the application of Theorem \ref{theorem:multivariate} in statistical contexts.

\section{Univariate autoregressive laws of motion}\label{sec:univariate}

We begin our discussion of autoregressive laws of motion with the univariate case, i.e.\ $N=1$. The results in this section are presented as three basic propositions. All three can be regarded as immediate corollaries to our Theorem \ref{theorem:multivariate} in Section \ref{sec:mainresult}. We do not prove the propositions but rather provide a brief commentary on each one; in any case all three can be proved directly using elementary methods, or deduced from discussions in prior literature. See, for instance, the treatment of difference equations in \citet[ch.~IX]{Sargent1987}. The point of commencing with the univariate case is to build intuition for the multivariate case, which is conceptually more challenging. The phrasing of the three propositions has been chosen to mimic the phrasing of Theorem \ref{theorem:multivariate}.

We take as given a real number $\phi$ which we call the autoregressive coefficient and a subexponential sequence of real numbers $\varepsilon=(\varepsilon_{t})$ which we call the innovation sequence. Our goal is to characterize the set of all sequences of real numbers $x=(x_t)$ that satisfy the infinite system of linear equations
\begin{equation}
	x_t=\phi x_{t-1}+\varepsilon_t,\quad t\in\mathbb Z.\label{eq:AR1}
\end{equation}
When \eqref{eq:AR1} is satisfied for a given sequence $x$ we say that $x$ is a solution to \eqref{eq:AR1}.

It will be useful to treat separately the cases where the magnitude of $\phi$ is less than, greater than, and equal to one. For reasons to become clear, we refer to \eqref{eq:AR1} as a \emph{forward} autoregressive law of motion if $\lvert\phi\rvert<1$, as a \emph{backward} autoregressive law of motion if $\lvert\phi\rvert>1$, and as an \emph{outward} autoregressive law of motion if $\lvert\phi\rvert=1$.

\subsection{Forward univariate autoregressive laws of motion}\label{sec:forward}

When $\phi=0$ the solution to \eqref{eq:AR1} is simply $x=\varepsilon$. This solution is subexponential and unique. Our first result characterizes the set of solutions to \eqref{eq:AR1} when $0<\lvert\phi\rvert<1$.

\begin{proposition}\label{theorem:AR1forward}
	Let $\phi$ be a real number satisfying $0<\lvert\phi\rvert<1$. Let $\varepsilon=(\varepsilon_t)$ be a subexponential sequence of real numbers. Let $x=(x_t)$ be a sequence of real numbers. The following two statements are equivalent.
	\begin{enumerate}[label=\upshape(\roman*)]
		\item $x_t=\phi x_{t-1}+\varepsilon_t$ for each $t\in\mathbb Z$, i.e.\ $x$ is a solution to \eqref{eq:AR1}.\label{en:forward-1}
		\item There exists $v\in\mathbb R$ such that $x_t=\phi^tv+\sum_{k=0}^\infty\phi^k\varepsilon_{t-k}$ for each $t\in\mathbb Z$.\label{en:forward-2}
		\setcounter{counter:items}{\value{enumi}}
	\end{enumerate}
	Moreover, if $x$ is a solution to \eqref{eq:AR1} then the following two statements are true.
	\begin{enumerate}[label=\upshape(\roman*)]
		\setcounter{enumi}{\value{counter:items}}
		\item The choice of $v$ in \ref{en:forward-2} is uniquely determined by the equality $v=\lim_{n\to\infty}\phi^nx_{-n}$.\label{en:forward-3}
		\item If $x$ is subexponential then $v=0$.\label{en:forward-4}
		\setcounter{counter:items}{\value{enumi}}
	\end{enumerate}
\end{proposition}

Proposition \ref{theorem:AR1forward} reveals that when $0<\lvert\phi\rvert<1$ there are infinitely many solutions to \eqref{eq:AR1}. In particular, the equality in statement \ref{en:forward-2} defines a one-to-one correspondence $v\leftrightarrow x$ between $\mathbb R$ and the set of solutions. When a given sequence $x$ solves \eqref{eq:AR1}, Proposition \ref{theorem:AR1forward} tells us that $x$ must satisfy
\begin{align*}
	x_t&=\phi^t\lim_{n\to\infty}\phi^nx_{-n}+\sum_{k=0}^\infty\phi^k\varepsilon_{t-k}\quad\text{for each }t\in\mathbb Z.
\end{align*}
The first and second terms are, respectively, the predetermined forward $x$-flow and the forward $\varepsilon$-flow in \eqref{eq:figure}. The other four flows are zero. The entries of the forward $\varepsilon$-flow are uniquely determined by current and past entries of $\varepsilon$ and are the same for every solution $x$, while the predetermined forward $x$-flow is different for each distinct solution $x$ and is uniquely determined by the real number $\lim_{n\to\infty}\phi^nx_{-n}$. This limit exists for every solution $x$. Restricting attention to any one solution amounts to placing an initial condition on $x$ by choosing the value of $\lim_{n\to\infty}\phi^nx_{-n}$. The particular initial condition $\lim_{n\to\infty}\phi^nx_{-n}=0$ yields the unique subexponential solution.

An obvious but important implication of the preceding discussion is that, when $0<\lvert\phi\rvert<1$, the autoregressive law of motion \eqref{eq:AR1} does not, on its own, uniquely determine a solution $x$. To obtain a unique solution one must also specify an initial condition for $x$ by choosing the value of $\lim_{n\to\infty}\phi^nx_{-n}$. Confining attention to the unique subexponential solution to \eqref{eq:AR1} is equivalent to imposing the initial condition $\lim_{n\to\infty}\phi^nx_{-n}=0$.

\subsection{Backward univariate autoregressive laws of motion}\label{sec:backward}

Our next result characterizes the set of solutions to \eqref{eq:AR1} when $\lvert\phi\rvert>1$.

\begin{proposition}\label{theorem:AR1backward}
	Let $\phi$ be a real number satisfying $\lvert\phi\rvert>1$. Let $\varepsilon=(\varepsilon_t)$ be a subexponential sequence of real numbers. Let $x=(x_t)$ be a sequence of real numbers. The following two statements are equivalent.
	\begin{enumerate}[label=\upshape(\roman*)]
		\item $x_t=\phi x_{t-1}+\varepsilon_t$ for each $t\in\mathbb Z$, i.e.\ $x$ is a solution to \eqref{eq:AR1}.\label{en:backward-1}
		\item There exists $v\in\mathbb R$ such that $x_t=\phi^tv-\sum_{k=1}^\infty\phi^{-k}\varepsilon_{t+k}$ for each $t\in\mathbb Z$.\label{en:backward-2}
		\setcounter{counter:items}{\value{enumi}}
	\end{enumerate}
	Moreover, if $x$ is a solution to \eqref{eq:AR1} then the following two statements are true.
	\begin{enumerate}[label=\upshape(\roman*)]
		\setcounter{enumi}{\value{counter:items}}
		\item The choice of $v$ in \ref{en:backward-2} is uniquely determined by the equality $v=\lim_{n\to\infty}\phi^{-n}x_{n}$.\label{en:backward-3}
		\item If $x$ is subexponential then $v=0$.\label{en:backward-4}
		\setcounter{counter:items}{\value{enumi}}
	\end{enumerate}
\end{proposition}

In the backward case with $\lvert\phi\rvert>1$ statement \ref{en:backward-2} defines a one-to-one correspondence $v\leftrightarrow x$ between $\mathbb R$ and the set of solutions. Thus there are infinitely many solutions when $\lvert\phi\rvert>1$. When a given sequence $x$ solves \eqref{eq:AR1}, Proposition \ref{theorem:AR1backward} tells us that $x$ must satisfy
\begin{equation}
	x_t=\phi^t\lim_{n\to\infty}\phi^{-n}x_{n}-\sum_{k=1}^\infty\phi^{-k}\varepsilon_{t+k}\quad\text{for each }t\in\mathbb Z.\label{eq:backwardrep}
\end{equation}
The first and second terms are, respectively, the predetermined backward $x$-flow and the backward $\varepsilon$-flow in \eqref{eq:figure}. The other four flows are zero. The entries of the backward $\varepsilon$-flow are uniquely determined by future entries of $\varepsilon$ and are the same for every solution $x$, while the predetermined backward $x$-flow is different for each distinct solution $x$ and is uniquely determined by the real number $\lim_{n\to\infty}\phi^{-n}x_n$. This limit exists for every solution $x$. One may single out a particular solution $x$ by choosing the value of $\lim_{n\to\infty}\phi^{-n}x_n$. This choice may be viewed as an initial condition for $x$, with the understanding that, for a backward law of motion, initialization occurs in the arbitrarily distant future. The future initialization $\lim_{n\to\infty}\phi^{-n}x_n=0$ yields the unique subexponential solution.

The decomposition of $x$ provided in \eqref{eq:backwardrep} is closely related to recent literature on asset price bubbles. \citet{HiranoToda2025a} consider an infinite-horizon deterministic economy in which an asset pays dividend $d_t$ and trades at ex-dividend price $p_t$ at each time $t\in\mathbb N\cup\{0\}$. Under standard conditions including no-arbitrage it is shown that
\begin{equation}
	p_t=\frac{1}{q_t}\lim_{n\to\infty}q_np_n+\frac{1}{q_t}\sum_{k=1}^\infty q_{t+k}d_{t+k}\quad\text{for each }t\in\mathbb N\cup\{0\},\label{eq:assetpricing}
\end{equation}
where $q_0=1$ and $q_1,q_2,\dots$ is a sequence of Arrow-Debreu prices for future delivery of one unit of the asset paid at time zero. See Eq.\ 6 in \citet[p.~116]{HiranoToda2025a}. If we set $q_t=\phi^{-t}$, $p_t=x_t$ and $d_t=-\varepsilon_t$ in \eqref{eq:assetpricing} then we recover the equality in \eqref{eq:backwardrep}. Hirano and Toda refer to the first term on the right-hand side of the equality in \eqref{eq:assetpricing} as the \emph{bubble component} of the asset, and refer to the second term as the \emph{fundamental value} of the asset. They refer to the equality $\lim_{n\to\infty}q_np_n=0$ as the \emph{transversality condition for asset pricing} and argue that this condition necessarily fails to hold in plausible model economies, leading to the presence of asset price bubbles. See also Eqs.\ 2.5.9--14 in \citet[pp.~38--9]{Hamilton1994}, Eqs.\ 20.8 and 20.13 in \citet[pp.~398--402]{Cochrane2005} and Eqs.\ 2--4 in \citet[pp.~3--4]{HiranoToda2025b}. By adapting the flow terminology introduced in this article we may say that the fundamental value of the asset is its backward dividend-flow, that the bubble component of the asset is its predetermined backward price-flow, and that the transversality condition for asset pricing is satisfied when the predetermined backward price-flow is zero.

\subsection{Outward univariate autoregressive laws of motion}\label{sec:outward}

Our final result in this section characterizes the set of solutions to \eqref{eq:AR1} when $\lvert\phi\rvert=1$; that is, when $\phi\in\{-1,1\}$.

\begin{proposition}\label{theorem:AR1outward}
	Let $\phi\in\{-1,1\}$. Let $\varepsilon=(\varepsilon_t)$ be a subexponential sequence of real numbers. Let $x=(x_t)$ be a sequence of real numbers. The following two statements are equivalent.
	\begin{enumerate}[label=\upshape(\roman*)]
		\item $x_t=\phi x_{t-1}+\varepsilon_t$ for each $t\in\mathbb Z$, i.e.\ $x$ is a solution to \eqref{eq:AR1}.\label{en:outward-1}
		\item There exists $v\in\mathbb R$ such that\label{en:outward-2}
		\begin{equation*}
			x_t=\phi^tv+\begin{cases}-\sum_{s=0}^{-t-1}\phi^{t+s}\varepsilon_{-s}&\text{for each negative }t\in\mathbb Z\\
				0&\text{for }t=0\\
				\sum_{s=1}^t\phi^{t-s}\varepsilon_s&\text{for each positive }t\in\mathbb Z.\end{cases}
		\end{equation*}
	\end{enumerate}
	Moreover, if $x$ is a solution to \eqref{eq:AR1} then the following two statements are true.
	\begin{enumerate}[label=\upshape(\roman*)]
		\setcounter{enumi}{\value{counter:items}}
		\item The choice of $v$ in \ref{en:outward-2} is uniquely determined by the equality $v=x_0$.\label{en:outward-3}
		\item $x$ is subexponential.\label{en:outward-4}
		\setcounter{counter:items}{\value{enumi}}
	\end{enumerate}
\end{proposition}

Similar to Propositions \ref{theorem:AR1forward} and \ref{theorem:AR1backward}, we see from Proposition \ref{theorem:AR1outward} that when $\lvert\phi\rvert=1$ there is a one-to-one correspondence $v\leftrightarrow x$ between $\mathbb R$ and the set of solutions. Thus, in all cases with $\phi\neq0$, there are infinitely many solutions. When $\lvert\phi\rvert=1$ and a given sequence $x$ solves \eqref{eq:AR1}, Proposition \ref{theorem:AR1outward} tells us that $x$ must satisfy
\begin{equation*}
	x_t=\phi^tx_0+\begin{cases}-\sum_{s=0}^{-t-1}\phi^{t+s}\varepsilon_{-s}&\text{for each negative }t\in\mathbb Z\\
		0&\text{for }t=0\\
		\sum_{s=1}^t\phi^{t-s}\varepsilon_s&\text{for each positive }t\in\mathbb Z.\end{cases}
\end{equation*}
The first term on the right-hand side of the last equality is the predetermined outward $x$-flow in \eqref{eq:figure}. The second term, written separately for negative, zero and positive $t\in\mathbb Z$, is the outward $\varepsilon$-flow. The other four flows are zero.

Note the outward character of the $\varepsilon$-flow with $\varepsilon_1\to\phi\varepsilon_1+\varepsilon_2\to\phi^2\varepsilon_1+\phi\varepsilon_2+\varepsilon_3$ and so forth as we move forward in time and $-\phi^{-1}\varepsilon_0\to-\phi^{-2}\varepsilon_0-\phi^{-1}\varepsilon_{-1}\to-\phi^{-3}\varepsilon_0-\phi^{-2}\varepsilon_{-1}-\phi^{-1}\varepsilon_{-2}$ and so forth as we move backward in time. Thus, moving outward from time zero in either direction, we gradually accumulate innovations along the path traveled. The predetermined outward $x$-flow depends on $x$ only through $x_0$.

We learned in Sections \ref{sec:forward} and \ref{sec:backward} that when $\lvert\phi\rvert\neq1$ there is exactly one subexponential solution to \eqref{eq:AR1}. This is not true when $\lvert\phi\rvert=1$: Proposition \ref{theorem:AR1outward} establishes that, in this case, all of the infinitely many solutions to \eqref{eq:AR1} are subexponential.

\section{Multivariate autoregressive laws of motion}\label{sec:multivariate}

We now turn to the more general multivariate case with $N\in\mathbb N$ variables. Throughout this section we fix a real $N\times N$ matrix $\Phi$ and a subexponential sequence of real $N\times1$ vectors $\varepsilon=(\varepsilon_t)$. Our goal is to characterize the set of all sequences of real $N\times 1$ vectors $x=(x_t)$ that are solutions to \eqref{eq:VAR1}.

Unlike the univariate case, we do not necessarily regard a multivariate autoregressive law of motion to have an exclusively forward, backward, or outward character. Depending on the eigenvalues of $\Phi$, all three directions may be relevant. The approach we will pursue involves using spectral projection to separate each solution into three components, corresponding respectively to eigenvalues of $\Phi$ which are inside the unit circle, outside the unit circle, and on the unit circle. We will see that the laws of motion for the three components of $x$ are similar to the respective forward, backward and outward univariate autoregressive laws of motion studied in Section \ref{sec:univariate}.  The construction of spectral projections from $\Phi$ is discussed in Section \ref{sec:spectralprojection}. The representations we provide for the forward, backward and outward components of each solution involve the Drazin inverse of $\Phi$. We discuss the Drazin inverse in Section \ref{sec:Drazin}. Finally we present our main result characterizing the set of all solutions in Section \ref{sec:mainresult}.

\subsection{Spectral projection}\label{sec:spectralprojection}

Let $\sigma$ denote the spectrum of $\Phi$; that is, the set of all complex eigenvalues of $\Phi$. Recall that a complex $N\times 1$ vector $v$ is called a \emph{generalized eigenvector} of $\Phi$ associated with the eigenvalue $\lambda\in\sigma$ if $(\Phi-\lambda I)^Nv=0$. The subspace of $\mathbb C^N$ spanned by all generalized eigenvectors of $\Phi$ associated with $\lambda$ is called the \emph{generalized eigenspace} of $\Phi$ associated with $\lambda$. The dimension of each generalized eigenspace is equal to the algebraic multiplicity of the corresponding eigenvalue of $\Phi$.

To each subset of eigenvalues $A\subseteq\sigma$ there corresponds a unique $N\times N$ projection matrix $P_A$ called a \emph{spectral projection}. If $A$ is empty then we define $P_A$ to be the $N\times N$ zero matrix, while if $A=\sigma$ then we define $P_A$ to be the $N\times N$ identity matrix. Otherwise we define $P_A$ as follows. Let $m_A$ be the sum of the algebraic multiplicities of the eigenvalues in $A$. If $A$ is a nonempty strict subset of $\sigma$ then $1\leq m_A<N$. Let $V_A$ be a complex $N\times m_A$ matrix whose column space is the direct sum of the generalized eigenspaces of $\Phi$ associated with the eigenvalues in $A$. Let $W_A$ be a complex $N\times(N-m_A)$ matrix whose column space is the direct sum of the other generalized eigenspaces of $\Phi$. Let $W_A^\perp$ be a full rank complex $m_A\times N$ matrix whose rows are orthogonal to the columns of $W_A$. The spectral projection $P_A$ is the complex $N\times N$ matrix defined by the equality
\begin{equation*}
	P_A=V_A(W_A^\perp V_A)^{-1}W_A^\perp.
\end{equation*}
Thus $P_A$ is the projection on the column space of $V_A$ along the column space of $W_A$. See, for instance, \citet[p.~168]{BanerjeeRoy2014}. Since the column spaces of $V_A$ and of $W_A$ are uniquely determined by $\Phi$ and $A$, the spectral projection $P_A$ is also uniquely determined by $\Phi$ and $A$.

In practice a Jordan decomposition of $\Phi$ may be used to choose the matrices $V_A$ and $W_A$. Let $J$ be the Jordan normal form of $\Phi$, so that $\Phi=VJV^{-1}$ for some nonsingular complex $N\times N$ matrix $V$. The eigenvalues of $\Phi$ lie on the diagonal of $J$, repeated according to algebraic multiplicity. We may choose the columns of $V_A$ to be the $m_A$ columns of $V$ corresponding to the diagonal entries of $J$ which belong to $A$, and choose the columns of $W_A$ to be the other $N-m_A$ columns of $V$.  That is, $V = [V_A \mid W_A]$. In this context an alternative formula for the $N \times N$ projection matrix is
\begin{equation*}
	P_A = [V_A \mid \bigzero ] V^{-1}.
\end{equation*}
Indeed the $m_A \times N$ matrix product $(W_A^\perp V_A)^{-1}W_A^\perp$ is simply the first $m_A$ rows of the inverse matrix $V^{-1}$.  Thus the two definitions are equivalent.  The alternative formula shows that orthogonality is not intrinsic to the projection itself but is simply an \emph{artefact} of the previous construction.

The next lemma lists commonly used properties of spectral projection matrices. These properties are well-known in the mathematical literature and may be deduced from, for instance, the discussion of generalized eigenspace decomposition in \citet[ch.~8]{Axler2024}. Recall that the \emph{index} of an eigenvalue $\lambda$ of $\Phi$ is the size of the largest Jordan block associated with $\lambda$ in the Jordan normal form of $\Phi$.
\begin{lemma}[Properties of spectral projections]\label{lemma:spectralprojection}
	Let $\Phi$ be a real square matrix and let $\sigma$ be the spectrum of $\Phi$. If $A$ is a subset of $\sigma$ then:
	\begin{enumerate}[label=\upshape(\roman*)]
		\item $P_A^2=P_A$.\label{en:projection}
		\item $P_A\Phi=\Phi P_A$.\label{en:commute}
		\item The spectrum of $\Phi P_A$ is equal to $A\cup\{0\}$ if $A\neq\sigma$, or else equal to $A$ if $A=\sigma$.\label{en:spectrum}
		\setcounter{counter:items}{\value{enumi}}
	\end{enumerate}
	If $A_1$ and $A_2$ are disjoint subsets of $\sigma$ then:
	\begin{enumerate}[label=\upshape(\roman*)]
		\setcounter{enumi}{\value{counter:items}}
		\item $P_{A_1}P_{A_2}=0$.\label{en:annihilate}
		\item $P_{A_1}+P_{A_2}=P_{A_1\cup A_2}$.\label{en:resolution}
		\setcounter{counter:items}{\value{enumi}}
	\end{enumerate}
	If $\lambda$ is an eigenvalue of $\Phi$ then:
	\begin{enumerate}[label=\upshape(\roman*)]
		\setcounter{enumi}{\value{counter:items}}
		\item $(\Phi-\lambda I)P_{\{\lambda\}}$ is nilpotent, with degree of nilpotency equal to the index of $\lambda$.\label{en:nilpotent}
		\setcounter{counter:items}{\value{enumi}}
	\end{enumerate}
	If $A$ is a subset of $\sigma$ then $\bar{A}$ is also a subset of $\sigma$ and:
	\begin{enumerate}[label=\upshape(\roman*)]
		\setcounter{enumi}{\value{counter:items}}
		\item $P_{\bar{A}} = \bar{P}_A$.\label{en:real}
		\setcounter{counter:items}{\value{enumi}}
	\end{enumerate}
\end{lemma}
For the final property we note with respect to the above partition $V = [V_A \mid W_A]$ that the columns of $\bar{V}_A$ and $\bar{W}_A$ are the respective generalized eigenvectors for the eigenvalues of $\Phi$ that lie in $\bar{A}$ and the eigenvalues of $\Phi$ that do not lie in $\bar{A}$. Thus
\begin{equation*}
	P_{\bar{A}} = [\bar{V}_A \mid \bigzero]\, {\rule{0cm}{0.3cm}\bar{V}}^{\,-1} = \overline{[V_A \mid \bigzero] V^{-1}}=\bar{P}_A
\end{equation*}
because the inverse of the conjugate of $V$ is the conjugate of the inverse of $V$.  It follows that $P_A + P_{\bar{A}}$ is real and that $P_A - P_{\bar{A}}$ is pure imaginary.

We adopt a special notation for the spectral projections and associated column spaces used in our main result characterizing the various directional flows in the components of our solutions to the corresponding multivariate autoregressive law of motion. Each such projection $P_A=[V_A\mid\bigzero]V^{-1}$ is a real $N\times N$ matrix with a real column space $\mathbb V_A=\col(V_A)=\col(P_A)\subseteq\mathbb R^N$. Table \ref{table:notation} shows the notation to be used. The projections defined in Table \ref{table:notation} are $P_\bullet$, $P_{\dotrightarrow}$, $P_\rightarrow$, $P_\leftarrow$ and $P_\leftrightarrow$. The corresponding subspaces are $\mathbb V_\bullet$, $\mathbb V_{\dotrightarrow}$, $\mathbb V_\rightarrow$, $\mathbb V_\leftarrow$ and $\mathbb V_\leftrightarrow$. The arrow subscripts are intended to suggest a directional classification---forward, backward or outward---for the eigenvalues in the key subsets $A\subseteq\sigma$ and the associated component flows. We elaborate further upon the concept of a directed flow in Section \ref{sec:arrow}. Note that $P_{\dotrightarrow}+P_\leftarrow+P_\leftrightarrow=I$ and $P_{\dotrightarrow}=P_\bullet+P_\rightarrow$.

\begin{table*}
	\caption{Notation for spectral projections and their column spaces.}
	\label{table:notation}
	\begin{center}
		\begin{tabular}{@{}lcc@{}@{}}
			\hline
			Spectral subset $A$
			& Notation for $P_A$
			& Notation for $\col(P_A)$ \\
			\hline
			$A=\{\lambda\in\sigma:\lambda=0\}$    & $P_\bullet$ & $\mathbb V_\bullet$ \\
			$A=\{\lambda\in\sigma:\lvert\lambda\rvert<1\}$    & $P_{\dotrightarrow}$    & $\mathbb V_{\dotrightarrow}$ \\
			$A=\{\lambda\in\sigma:0<\lvert\lambda\rvert<1\}$    & $P_\rightarrow$    & $\mathbb V_\rightarrow$ \\
			$A=\{\lambda\in\sigma:\lvert\lambda\rvert>1\}$    & $P_\leftarrow$    & $\mathbb V_\leftarrow$ \\
			$A=\{\lambda\in\sigma:\lvert\lambda\rvert=1\}$    & $P_\leftrightarrow$    & $\mathbb V_\leftrightarrow$ \\
			\hline
		\end{tabular}
	\end{center}
	\small{The second column shows the special notation used for the spectral projection $P_A$ when $A$ is the subset of the spectrum of $\Phi$ shown in the first column. All spectral subsets $A$ shown in the first column are closed under complex conjugation because $\Phi$ is a real matrix. Thus all of the corresponding spectral projections are real matrices. The third column shows the notation used for the real column spaces of these spectral projections.}
\end{table*}

\subsection{The Drazin inverse}\label{sec:Drazin}

The characterization of the forward and backward components of the solutions to a multivariate autoregressive law of motion supplied by our main result in Section \ref{sec:mainresult} makes use of a particular generalized inverse of $\Phi$ called the \emph{Drazin inverse}. The Drazin inverse of a complex $N\times N$ matrix $M$, denoted $M^\mathrm{D}$, is the unique complex $N\times N$ matrix that satisfies
\begin{align}
	M^\mathrm{D}MM^\mathrm{D}&=M^\mathrm{D},\label{eq:Drazin1}\\
	M^\mathrm{D}M&=MM^\mathrm{D},\label{eq:Drazin2}\\
	\text{and}\quad M^\mathrm{D}M^{N+1}&=M^N.\label{eq:Drazin3}
\end{align}
The fact that these three properties uniquely define $M^\mathrm{D}$ is shown in, for instance, \citet[ch.~7]{CampbellMeyer1979}. If all entries of $M$ are real then all entries of $M^\mathrm{D}$ are also real. We say, informally, that $M^\mathrm{D}$ is a generalized inverse of $M$ because it coincides with the ordinary inverse $M^{-1}$ if $M$ is nonsingular, and more generally satisfies the inverse-like properties \eqref{eq:Drazin1}--\eqref{eq:Drazin3}. It should be noted, however, that $M^\mathrm{D}$ does not in general satisfy $MM^\mathrm{D}M=M$, and so is not a generalized inverse in the strict sense in which some authors use this term.

The Drazin inverse is easily constructed from the Jordan normal form. When $\Phi$ is nonsingular $\Phi^\mathrm{D}$ is simply the usual inverse matrix $\Phi^{-1}$. Now suppose $\Phi$ is singular. Let $V$ be a nonsingular complex $N\times N$ matrix such that $J=V^{-1}\Phi V$ is the Jordan normal form of $\Phi$, with eigenvalues ordered along the main diagonal of $J$ in such a way that
\begin{equation*}
	J=\begin{bmatrix}J_R&\bigzero\\\bigzero&J_S\end{bmatrix}
\end{equation*}
where all diagonal elements of $J_R$ are nonzero and all diagonal elements of $J_S$ are zero. We have $\Phi=VJV^{-1}$. The Drazin inverse of $\Phi$ is defined by
\begin{equation*}
	\Phi^\mathrm{D} = V \begin{bmatrix}J_R^{-1} & \bigzero \\
		\bigzero & \bigzero \end{bmatrix} V^{-1}.\label{eq:DrazinJordan}
\end{equation*}

The following lemma states several convenient properties of the Drazin inverse related to the spectral projection matrices introduced in Section \ref{sec:spectralprojection}.

\begin{lemma}[Spectral projections and the Drazin inverse]\label{lemma:Drazin}
	Let $\Phi$ be a real square matrix and let $\sigma$ be the spectrum of $\Phi$. If $A$ is a subset of $\sigma$ then:
	\begin{enumerate}[label=\upshape(\roman*)]
		\item $P_A^\mathrm{D}=P_A$.\label{en:projectionDrazin}
		\item $P_A\Phi^\mathrm{D}=\Phi^\mathrm{D} P_A$.\label{en:commuteDrazin}
		\item The spectrum of $\Phi^\mathrm{D} P_A$ is the set of all nonzero $\mu\in\mathbb C$ such that $\mu^{-1}\in A$, together with the point $\mu=0$ if either $0\in A$ or $A\subset\sigma$.\label{en:spectrumDrazin}
		\item If $0\notin A$ then $\Phi^\mathrm{D}\Phi P_A=P_A$.\label{en:restrictDrazin}
		\setcounter{counter:items}{\value{enumi}}
	\end{enumerate}
	If $\lambda$ is a nonzero eigenvalue of $\Phi$ with index one then:
	\begin{enumerate}[label=\upshape(\roman*)]
		\setcounter{enumi}{\value{counter:items}}
		\item $\Phi^\mathrm{D}P_{\{\lambda\}}=\lambda^{-1}P_{\{\lambda\}}$.\label{en:index1Drazin}
	\end{enumerate}
\end{lemma}

Lemma \ref{lemma:Drazin} follows from the properties of spectral projections stated in Lemma \ref{lemma:spectralprojection} and known properties of the Drazin inverse.  See \cite{CampbellMeyer1979}.  In particular we refer readers to Corollary 7.2.1 and Theorems 7.2.2, 7.4.1 and 7.8.4.

The Drazin inverse was developed in \cite{Drazin1958} and applied to problems involving linear systems of differential or difference equations in \cite{CampbellMeyerRose1976} and \cite{Campbell1979}. It appears infrequently in prior econometric literature. Examples include \cite{Neusser2000} and \cite{Zoia2009}, both concerning autoregressive laws of motion with a unit eigenvalue. Applications of the Drazin inverse to problems involving Markov chains and optimal control are discussed in \citet[ch.~8,9]{CampbellMeyer1979}.

\subsection{Main result}\label{sec:mainresult}

In this section the notation $\Phi^{-k}$ with $k\in\mathbb N$ refers to the $k$th power of the Drazin inverse of $\Phi$. At no point do we assume that $\Phi$ is invertible.

Our main result describes the general solution to a multivariate autoregressive law of motion.  In order to do this we use the frequency-specific difference and cumulation operators introduced in \cite{Gregoir1999a}.  Although our solution space is confined to sequences of real vectors, working with the frequency-specific difference and cumulation operators will entail the consideration of sequences of complex vectors. Let ${\mathbb S}$ be the linear space of all sequences $x = (x_t)$ of vectors $x_t \in {\mathbb C}^N$ indexed by the integer times $t \in {\mathbb Z}$. Addition and scalar multiplication in $\mathbb S$ are defined respectively by $(x + y)_t = x_t + y_t$ and $(cx)_t = cx_t$ for all complex numbers $c$ and all times $t \in {\mathbb Z}$.

In this context the familiar \emph{backshift operator} $B:{\mathbb S}\to{\mathbb S}$ is a linear operator defined by the formula
\begin{equation*}
	(Bx)_t=x_{t-1}\quad\text{for each }t\in\mathbb Z.
\end{equation*}
Although the backshift operator is not a matrix operator it is nevertheless fundamental to time series analysis.  For each value of the parameter $\theta\in(-\pi,\pi]$ we can now define the \emph{difference operator at frequency} $\theta$ as the linear map $D_\theta:{\mathbb S}\to{\mathbb S}$ given by
\begin{equation*}
	(D_\theta x)_t = x_t - e^{-i\theta}(Bx)_t \quad\text{for each }t\in\mathbb Z,
\end{equation*}
the \emph{cumulation operator at frequency} $\theta$ as the linear map $C_\theta:{\mathbb S}\to{\mathbb S}$ given by
\begin{equation*}
	(C_\theta x)_t=\begin{dcases}-\sum_{s=0}^{-t-1}e^{-i\theta(t+s)}x_{-s}&\text{for each negative }t\in\mathbb Z\\0&\text{for }t=0\\\sum_{s=1}^te^{-i\theta(t-s)}x_s&\text{for each positive }t\in\mathbb Z.\end{dcases}
\end{equation*}
and the \emph{residual operator at frequency} $\theta$ as the linear map $R_\theta:\mathbb C^N\to{\mathbb S}$ given by
\begin{equation*}
	(R_\theta v)_t = e^{-i\theta t}v \quad\text{for each }t\in\mathbb Z.
\end{equation*}
We note that in the final definition $v \in \mathbb C^N$ is a vector and $R_{\theta}v \in {\mathbb S}$ is a sequence whose value at the specific time $t$ is the vector $e^{-i\theta t}v \in \mathbb C^N$.

The frequency-specific cumulation and residual operators arise naturally in the solution of autoregressive laws of motion when we have one or more eigenvalues on the unit circle. The following lemma, adapted from \cite{Gregoir1999a}, establishes that the difference operator at frequency $\theta$ is a left-inverse of the cumulation operator at frequency $\theta$, but not a right-inverse. It also explains why we call $R_\theta$ the residual operator.

\begin{lemma}\label{lemma:gregoir}
	For each $\theta\in(-\pi,\pi]$ and each $x\in{\mathbb S}$ we have $D_\theta C_\theta x = x$, $C_\theta D_\theta x = x-R_\theta x_0$ and $D_\theta R_\theta x_0=0$, which we may rewrite entry-wise as $(D_\theta C_\theta x)_t = x_t$, $(C_\theta D_\theta x)_t = x_t-(R_\theta x_0)_t$ and $(D_\theta R_\theta x_0)_t=0$ for all $t \in {\mathbb Z}$.
\end{lemma}

See pp.~437--41 in \cite{Gregoir1999a} for a more detailed discussion of frequency-specific differencing and cumulation. In what follows we will make use of the notation
\begin{equation*}
	\Theta=\{\theta\in(-\pi,\pi]:e^{-i\theta}\in\sigma\}
\end{equation*}
where $\sigma$ is the spectrum of the autoregressive coefficient $\Phi$. The set $\Theta$ contains the frequencies of the eigenvalues of $\Phi$ on the unit circle. The matrix $\Phi$ is assumed to be real, so its spectrum $\sigma$ is closed under complex conjugation. Thus if $\theta\in\Theta$ and $\theta\neq\pi$ then $-\theta\in\Theta$. We will use the notation $P_\theta$ as shorthand for the spectral projection $P_{\{e^{-i\theta}\}}$ associated with the specific eigenvalue $e^{-i\theta}$, and we will write $d_\theta$ for the index of the eigenvalue $e^{-i\theta}$. Note that $P_\theta$ is a real matrix for $\theta\in\{0,\pi\}$, and that $P_\theta+P_{-\theta}$ is a real matrix for $\theta\in(0,\pi)$, by Lemma \ref{lemma:spectralprojection}. Also note that $P_\leftrightarrow=\sum_{\theta\in\Theta}P_\theta$ by Lemma \ref{lemma:spectralprojection}. We have $d_\theta=d_{-\theta}$ for $\theta\neq\pi$ because conjugate eigenvalues of a real matrix have the same index.

\begin{theorem}\label{theorem:multivariate}
	Let \(\Phi\) be a real $N\times N$ matrix with spectrum $\sigma$. Let $\varepsilon=(\varepsilon_t)$ be a subexponential sequence of real $N\times 1$ vectors. Let $x=(x_t)$ be a sequence of real $N\times 1$ vectors. Then $x_t=P_{\dotrightarrow}x_t+P_\leftarrow x_t+P_\leftrightarrow x_t$ for each $t\in\mathbb Z$, and the series $\sum_{k=0}^\infty\Phi^kP_{\dotrightarrow}\varepsilon_{t-k}$ and $\sum_{k=1}^\infty\Phi^{-k}P_\leftarrow\varepsilon_{t+k}$ converge for each $t\in\mathbb Z$. The following two statements are equivalent.
	\begin{enumerate}[label=\upshape(\roman*)]
		\item $x_t=\Phi x_{t-1}+\varepsilon_t$ for each $t\in\mathbb Z$, i.e.\ $x$ is a solution to \eqref{eq:VAR1}.\label{en:statement1}
		\item There exist real vectors $v_\rightarrow\in\mathbb V_\rightarrow$, $v_\leftarrow\in\mathbb V_\leftarrow$ and $v_\leftrightarrow\in\mathbb V_\leftrightarrow$ such that, for each $t\in\mathbb Z$,\label{en:statement2}
		\begin{align}
			P_{\dotrightarrow} x_t&=\Phi^tv_\rightarrow+\sum_{k=0}^\infty\Phi^k P_{\dotrightarrow}\varepsilon_{t-k},\label{eq:mainthm1}\\
			P_\leftarrow x_t&=\Phi^tv_\leftarrow-\sum_{k=1}^\infty\Phi^{-k}P_\leftarrow\varepsilon_{t+k},\label{eq:mainthm2}\\
			\text{and}\quad \quad P_\leftrightarrow x_t&=\sum_{\theta\in\Theta}\sum_{k=1}^{d_\theta}(\Phi-e^{-i\theta}I)^{k-1}P_\theta(C_\theta B)^{k-1}[(R_\theta v_\leftrightarrow)_t+(C_\theta\varepsilon)_t].\label{eq:mainthm3}
		\end{align}
		\setcounter{counter:items}{\value{enumi}}
	\end{enumerate}
	Moreover, if $x$ is a solution to \eqref{eq:VAR1} then the following two statements are true.
	\begin{enumerate}[label=\upshape(\roman*)]
		\setcounter{enumi}{\value{counter:items}}
		\item The choices of $v_\rightarrow\in\mathbb V_\rightarrow$, $v_\leftarrow\in\mathbb V_\leftarrow$ and $v_\leftrightarrow\in\mathbb V_\leftrightarrow$ in \ref{en:statement2} are uniquely determined by the equalities\label{en:statement3}
		\begin{equation*}
			v_\rightarrow=\lim_{n\to\infty}\Phi^n P_\rightarrow x_{-n},\quad v_\leftarrow=\lim_{n\to\infty}\Phi^{-n} P_\leftarrow x_{n}\quad\text{and}\quad v_\leftrightarrow=P_\leftrightarrow x_0.\label{eq:mainthmextra}
		\end{equation*}
		\item If $x$ is subexponential then $v_\rightarrow=0$ and $v_\leftarrow=0$.\label{en:statement4}
		\setcounter{counter:items}{\value{enumi}}
	\end{enumerate}
\end{theorem}
\begin{proof}
	The first assertion follows from the identity $P_{\dotrightarrow}+P_\leftarrow+P_\leftrightarrow=I$. We now show that the series $\sum_{k=0}^\infty\Phi^kP_{\dotrightarrow}\varepsilon_{t-k}$ and $\sum_{k=1}^\infty\Phi^{-k}P_\leftarrow\varepsilon_{t+k}$ converge for each $t\in\mathbb Z$. We have
	\begin{equation*}
		\left\Vert\sum_{k=m}^{m+n}\Phi^kP_{\dotrightarrow}\varepsilon_{t-k}\right\Vert\leq\sum_{k=m}^{m+n}\lVert\Phi^kP_{\dotrightarrow}\varepsilon_{t-k}\rVert\leq\sum_{k=m}^\infty\lVert\Phi^kP_{\dotrightarrow}\varepsilon_{t-k}\rVert=\sum_{k=m}^\infty\lVert(\Phi P_{\dotrightarrow})^k\varepsilon_{t-k}\rVert
	\end{equation*}
	for each $m,n\in\mathbb N$ because Lemma \ref{lemma:spectralprojection} shows that $P_{\dotrightarrow}$ is idempotent  and commutes with $\Phi$. Lemma \ref{lemma:spectralprojection} also shows that $\Phi P_{\dotrightarrow}$ has all eigenvalues inside the unit circle, and so we deduce from Gelfand's formula\footnote{For any $N\times N$ matrix $M$ and any $\delta>0$ we have $\lVert M^nv\rVert\leq(\rho(M)+\delta)^n\lVert v\rVert$ for all $N\times 1$ vectors $v$ and all sufficiently large $n\in\mathbb N$ where $\rho(M)$ is the spectral radius of $M$. See \citet[p.~497]{BanerjeeRoy2014}.} and the fact that $\varepsilon$ is subexponential that the last sum over $k$ vanishes in the limit as $m\to\infty$. Thus $\sum_{k=0}^\infty\Phi^kP_{\dotrightarrow}\varepsilon_{t-k}$ converges, by Cauchy's criterion. We similarly have
	\begin{equation*}
		\left\Vert\sum_{k=m}^{m+n}\Phi^{-k}P_\leftarrow\varepsilon_{t+k}\right\Vert\leq\sum_{k=m}^\infty\lVert\Phi^{-k}P_\leftarrow\varepsilon_{t+k}\rVert=\sum_{k=m}^\infty\lVert(\Phi^{\mathrm{D}} P_\leftarrow)^k\varepsilon_{t+k}\rVert
	\end{equation*}
	for each $m,n\in\mathbb N$ because Lemma \ref{lemma:spectralprojection} shows that $P_\leftarrow$ is idempotent and Lemma \ref{lemma:Drazin} shows that $P_\leftarrow$ commutes with $\Phi^\mathrm{D}$. Lemma \ref{lemma:Drazin} also shows that $\Phi^{\mathrm D} P_\leftarrow$ has all eigenvalues inside the unit circle. We note too that $\varepsilon$ is subexponential. Now it follows from Gelfand's formula that the final sum vanishes in the limit as $m\to\infty$.
	
	We next show that \ref{en:statement1} and \ref{en:statement2} are equivalent and that \ref{en:statement1} implies \ref{en:statement3}. First suppose that \ref{en:statement1} is true. We aim to show that \ref{en:statement2} and \ref{en:statement3} are true. First we show that \eqref{eq:mainthm1} is true for exactly one choice of $v_\rightarrow\in\mathbb V_\rightarrow$, this being $v_\rightarrow=\lim_{n\to\infty}\Phi^nP_\rightarrow x_{-n}$. Care is needed to deal with the possibility that zero is an eigenvalue of $\Phi$. Apply $P_{\dotrightarrow}$ to both sides of the equation $x_t=\Phi x_{t-1}+\varepsilon_t$ and use the fact that $P_{\dotrightarrow}$ and $\Phi$ commute to obtain
	\begin{equation*}
		P_{\dotrightarrow} x_t=\Phi P_{\dotrightarrow} x_{t-1}+P_{\dotrightarrow}\varepsilon_t\quad\text{for each }t\in\mathbb Z.
	\end{equation*}
	By iterating the last equation backward we obtain
	\begin{equation}
		P_{\dotrightarrow} x_t=\Phi^n P_{\dotrightarrow} x_{t-n}+\sum_{k=0}^{n-1}\Phi^k P_{\dotrightarrow}\varepsilon_{t-k}\quad\text{for each }n\in\mathbb N\text{ and each }t\in\mathbb Z.\label{eq:mainthmprf00}
	\end{equation}
	We know that $P_{\dotrightarrow}=P_\bullet+P_\rightarrow$ and $\Phi^nP_\bullet=(\Phi P_\bullet)^n$ for each $n\in\mathbb N$. Hence it follows that
	\begin{equation*}
		P_{\dotrightarrow} x_t=(\Phi P_{\bullet})^n x_{t-n}+\Phi^n P_\rightarrow x_{t-n}+\sum_{k=0}^{n-1}\Phi^k P_{\dotrightarrow}\varepsilon_{t-k}\quad\text{for each }n\in\mathbb N\text{ and each }t\in\mathbb Z.
	\end{equation*}
	The leading term $(\Phi P_\bullet)^n x_{t-n}$ is zero for $n\geq N$ because Lemma \ref{lemma:spectralprojection} shows that $\Phi P_\bullet$ has spectrum equal to $\{0\}$ and is consequently nilpotent. Therefore
	\begin{equation}
		P_{\dotrightarrow} x_t=\Phi^n P_\rightarrow x_{t-n}+\sum_{k=0}^{n-1}\Phi^k P_{\dotrightarrow}\varepsilon_{t-k}\quad\text{for each }n\in\mathbb N\text{ with }n\geq N\text{ and each }t\in\mathbb Z.\label{eq:mainthmprf01}
	\end{equation}
	It follows from \eqref{eq:mainthmprf01} that $\Phi^nP_\rightarrow x_{t-n}\to P_{\dotrightarrow}x_t-\sum_{k=0}^\infty\Phi^k P_{\dotrightarrow}\varepsilon_{t-k}$ as $n\to\infty$ for each $t\in\mathbb Z$. The properties of the Drazin inverse show that $\Phi^n=\Phi^t\Phi^{n-t}$ for each $t\in\mathbb Z$ and each $n\in\mathbb N$ with $n\geq t$. Therefore
	\begin{equation*}
		\lim_{n\to\infty}\Phi^nP_\rightarrow x_{t-n}=\lim_{n\to\infty}\Phi^{t}\Phi^{n-t}P_\rightarrow x_{t-n}=\Phi^t\lim_{n\to\infty}\Phi^nP_\rightarrow x_{-n}\quad\text{for each }t\in\mathbb Z.
	\end{equation*}
	Taking the limit as $n\to\infty$ in \eqref{eq:mainthmprf01}, we find that \eqref{eq:mainthm1} is true for the particular choice of $v_\rightarrow\in\mathbb V_\rightarrow$ given by $v_\rightarrow=\lim_{n\to\infty}\Phi^nP_\rightarrow x_{-n}$. Note that $v_\rightarrow\in\mathbb{V}_\rightarrow$ because $\Phi^nP_\rightarrow x_{-n}=P_\rightarrow\Phi^nx_{-n}\in\col(P_\rightarrow)=\mathbb{V}_\rightarrow$ for each $n\in\mathbb N$. Setting $t=0$ in \eqref{eq:mainthm1} shows that the choice of $v_\rightarrow$ is unique.
	
	Next we show that \eqref{eq:mainthm2} is true for exactly one choice of $v_\leftarrow\in\mathbb V_\leftarrow$, this being $v_\leftarrow=\lim_{n\to\infty}\Phi^{-n}P_\leftarrow x_n$. Apply $\Phi^\mathrm{D}P_\leftarrow$ to both sides of the equation $x_{t+1}=\Phi x_{t}+\varepsilon_{t+1}$ and use the fact that $P_\leftarrow$ and $\Phi^\mathrm{D}$ commute and the identity $\Phi^\mathrm{D}\Phi P_\leftarrow=P_\leftarrow$ to obtain
	\begin{equation*}
		P_\leftarrow x_t=\Phi^\mathrm{D}P_\leftarrow x_{t+1}-\Phi^\mathrm{D}P_\leftarrow\varepsilon_{t+1}\quad\text{for each }t\in\mathbb Z.
	\end{equation*}
	Iterate the last equation forward to obtain
	\begin{equation}
		P_\leftarrow x_t=\Phi^{-n}P_\leftarrow x_{t+n}-\sum_{k=1}^n\Phi^{-k}P_\leftarrow\varepsilon_{t+k}\quad\text{for each }n\in\mathbb N\text{ and each }t\in\mathbb Z.\label{eq:mainthmprf02}
	\end{equation}
	It follows from \eqref{eq:mainthmprf02} that $\Phi^{-n}P_\leftarrow x_{t+n}\to P_{\leftarrow}x_t+\sum_{k=1}^\infty\Phi^{-k} P_{\leftarrow}\varepsilon_{t+k}$ as $n\to\infty$ for each $t\in\mathbb Z$. The properties of the Drazin inverse show that $\Phi^{-n}=\Phi^t\Phi^{-t-n}$ for each $t\in\mathbb Z$ and each $n\in\mathbb N$. Therefore
	\begin{equation*}
		\lim_{n\to\infty}\Phi^{-n}P_\leftarrow x_{t+n}=\lim_{n\to\infty}\Phi^t\Phi^{-t-n}P_\leftarrow x_{t+n}=\Phi^t\lim_{n\to\infty}\Phi^{-n}P_\leftarrow x_{n}\quad\text{for each }t\in\mathbb Z.
	\end{equation*}
	Taking the limit as $n\to\infty$ in \eqref{eq:mainthmprf02}, we find that \eqref{eq:mainthm2} is true for the particular choice of $v_\leftarrow\in\mathbb V_\leftarrow$ given by $v_\leftarrow=\lim_{n\to\infty}\Phi^{-n}P_\leftarrow x_{n}$. Setting $t=0$ in \eqref{eq:mainthm2} shows that the choice of $v_\leftarrow$ is unique.
	
	Next we show that \eqref{eq:mainthm3} is true for exactly one choice of $v_\leftrightarrow\in\mathbb V_\leftrightarrow$, this being $v_\leftrightarrow=P_\leftrightarrow x_0$. If $\Theta$ is empty then $\mathbb V_\leftrightarrow=\{0\}$ and $P_\leftrightarrow=\bigzero$, so the desired conclusion is trivially obtained. Suppose instead that $\Theta$ is nonempty. We will make use of the frequency-specific differencing and cumulation operators discussed earlier in this Section. Fix $\theta\in\Theta$ and subtract $e^{-i\theta}(Bx)_t$ from both sides of the equation $x_t=\Phi (Bx)_t+\varepsilon_t$ to obtain $(D_\theta x)_t=(\Phi-e^{-i\theta}I)(Bx)_t+\varepsilon_t$ for each $t\in\mathbb Z$. Apply $P_\theta C_\theta$ to both sides and use Lemma \ref{lemma:gregoir} and the fact that $P_\theta$ commutes with $\Phi$, $C_\theta$ and $B$ to show that
	\begin{equation}\label{eq:mainthmprf2}
		P_\theta x_t=P_\theta [(R_\theta x_0)_t+(C_\theta\varepsilon)_t]+(\Phi-e^{-i\theta}I)(C_\theta B)P_\theta x_t\quad\text{for each }t\in\mathbb Z.
	\end{equation}
	Observe that $P_\theta x_t$ appears on both sides of \eqref{eq:mainthmprf2}. If $d_\theta>1$ we substitute the entire right-hand side of \eqref{eq:mainthmprf2} in place of $P_{\theta}x_t$ on the right-hand side of \eqref{eq:mainthmprf2} to obtain
	\begin{multline*}
		P_{\theta}x_t=P_{\theta}[(R_\theta x_0)_t+(C_\theta\varepsilon)_t]+(\Phi-e^{-i\theta}I)P_{\theta}C_\theta B[(R_\theta x_0)_t+(C_\theta\varepsilon)_t]\\+(\Phi-e^{-i\theta}I)^2(C_\theta B)^2P_{\theta}x_t\quad\text{for each }t\in\mathbb Z.
	\end{multline*}
	After $d_\theta-1$ iterations of this substitution we obtain
	\begin{multline*}
		P_{\theta}x_t=\sum_{k=1}^{d_\theta}(\Phi-e^{-i\theta}I)^{k-1}P_{\theta}(C_\theta B)^{k-1}[(R_\theta x_0)_t+(C_\theta\varepsilon)_t]\\+(\Phi-e^{-i\theta}I)^{d_\theta}P_{\theta}(C_\theta B)^{d_\theta}x_t\quad\text{for each }t\in\mathbb Z,
	\end{multline*}
	which is the same as \eqref{eq:mainthmprf2} if $d_\theta=1$. The final term is zero because Lemma \ref{lemma:spectralprojection} shows that $(\Phi-e^{-i\theta}I)^{d_\theta}P_\theta=[(\Phi-e^{-i\theta}I)P_\theta]^{d_\theta}$ and that $(\Phi-e^{-i\theta}I)P_\theta$ is nilpotent of degree $d_\theta$. We can replace $x_0$ with $P_\leftrightarrow x_0$ in the summation over $k$ by noting that
	\begin{align*}
		P_\theta(C_\theta B)^{k-1}(R_\theta P_\leftrightarrow x_0)_t&=P_\theta P_\leftrightarrow(C_\theta B)^{k-1}(R_\theta x_0)_t\\
		&=P_\theta(C_\theta B)^{k-1}(R_\theta x_0)_t\quad\text{for each }t\in\mathbb Z
	\end{align*}
	because $P_\theta P_\leftrightarrow=P_\theta$ and $P_\leftrightarrow$ commutes with $C_\theta$, $R_\theta$ and $B$. Thus
	\begin{equation*}
		P_\theta x_t=\sum_{k=1}^{d_\theta}(\Phi-e^{-i\theta}I)^{k-1}P_\theta(C_\theta B)^{k-1}[(R_\theta P_\leftrightarrow x_0)_t+(C_\theta\varepsilon)_t]\quad\text{for each }t\in\mathbb Z.
	\end{equation*}
	Since $P_\leftrightarrow=\sum_{\theta\in\Theta}P_\theta$, by summing over all $\theta\in\Theta$ we deduce that \eqref{eq:mainthm3} is true for the particular choice of $v_\leftrightarrow\in\mathbb V_\leftrightarrow$ given by $v_\leftrightarrow=P_\leftrightarrow x_0$. To show that this choice of $v_\leftrightarrow\in\mathbb V_\leftrightarrow$ is unique simply substitute $t=0$ in \eqref{eq:mainthm3} and use the identity $(C_\theta y)_0=0$ for each $y\in\mathbb S$ and each $\theta\in\Theta$. Hence \eqref{eq:mainthm3} implies that
	\begin{equation*}
		P_\leftrightarrow x_0=\sum_{\theta\in\Theta}P_\theta (R_\theta v_\leftrightarrow)_0=\sum_{\theta\in\Theta}P_\theta v_\leftrightarrow=P_\leftrightarrow v_\leftrightarrow=v_\leftrightarrow.
	\end{equation*}
	
	We have shown that if \ref{en:statement1} is true then \ref{en:statement2} and \ref{en:statement3} are true. Next we show that if \ref{en:statement2} is true then \ref{en:statement1} is true. Suppose that \ref{en:statement2} is true. We will show that \ref{en:statement1} must be true by showing that $P_{\dotrightarrow} x_t=\Phi P_{\dotrightarrow} x_{t-1}+P_{\dotrightarrow}\varepsilon_t$ and $P_\leftarrow x_t=\Phi P_\leftarrow x_{t-1}+P_\leftarrow\varepsilon_t$ for each $t\in\mathbb Z$, and that $P_\theta x_t=\Phi P_\theta x_{t-1}+P_\theta\varepsilon_t$ for each $t\in\mathbb Z$ and each $\theta\in\Theta$. Summing the equalities yields \ref{en:statement1} because the spectral projections commute with $\Phi$ and because $P_{\dotrightarrow}+P_\leftarrow+\sum_{\theta\in\Theta}P_\theta=I$.
	
	To show that $P_{\dotrightarrow} x_t=\Phi P_{\dotrightarrow} x_{t-1}+P_{\dotrightarrow}\varepsilon_t$ for each $t\in\mathbb Z$ we use \eqref{eq:mainthm1} to write
	\begin{align*}
		\Phi P_{\dotrightarrow} x_{t-1}+P_{\dotrightarrow}\varepsilon_t&=\Phi\left(\Phi^{t-1}v_\rightarrow+\sum_{k=0}^\infty\Phi^kP_{\dotrightarrow}\varepsilon_{t-k-1}\right)+P_{\dotrightarrow}\varepsilon_t\\&=\Phi^tv_\rightarrow+\sum_{k=0}^\infty\Phi^kP_{\dotrightarrow}\varepsilon_{t-k}=P_{\dotrightarrow} x_t\quad\text{for each }t\in\mathbb Z.
	\end{align*}
	To show that $P_\leftarrow x_t=\Phi P_\leftarrow x_{t-1}+P_\leftarrow\varepsilon_t$ for each $t\in\mathbb Z$ we use \eqref{eq:mainthm2} to write
	\begin{align*}
		\Phi P_\leftarrow x_{t-1}+P_\leftarrow\varepsilon_t&=\Phi\left(\Phi^{t-1}v_\leftarrow-\sum_{k=1}^\infty\Phi^{-k}P_\leftarrow\varepsilon_{t+k-1}\right)+P_\leftarrow\varepsilon_t\\
		&=\Phi^tv_\leftarrow-\sum_{k=1}^\infty\Phi^{-k}P_\leftarrow\varepsilon_{t+k}=P_\leftarrow x_t\quad\text{for each }t\in\mathbb Z.
	\end{align*}
	To show that $P_\theta x_t=\Phi P_\theta x_{t-1}+P_\theta\varepsilon_t$ for each $t\in\mathbb Z$ we apply $P_\theta$ to both sides of \eqref{eq:mainthm3}. This yields
	\begin{equation}
		P_\theta x_t=\sum_{k=1}^{d_\theta}(\Phi-e^{-i\theta}I)^{k-1}P_\theta (C_\theta B)^{k-1}[(R_\theta v_\leftrightarrow)_t+(C_\theta\varepsilon)_t]\quad\text{for each }t\in\mathbb Z\label{eq:Ptheta}
	\end{equation}
	because $P_\theta$ commutes with $\Phi$ and because $P_\theta P_\leftrightarrow=P_\theta$ and $P_\theta P_{\theta'}=\bigzero$ for each $\theta,\theta'\in\Theta$ with $\theta'\neq\theta$. Next we apply $(\Phi-e^{-i\theta}I)B$ to both sides of \eqref{eq:Ptheta}, obtaining
	\begin{equation*}
		(\Phi-e^{-i\theta}I)P_\theta x_{t-1}=\sum_{k=1}^{d_\theta}(\Phi-e^{-i\theta}I)^{k}P_\theta B(C_\theta B)^{k-1}[(R_\theta v_\leftrightarrow)_t+(C_\theta\varepsilon)_t]\quad\text{for each }t\in\mathbb Z.
	\end{equation*}
	The final summand with $k=d_\theta$ is zero because $(\Phi-e^{-i\theta}I)^{d_\theta}P_\theta=[(\Phi-e^{-i\theta}I)P_\theta]^{d_\theta}$ and $(\Phi-e^{-i\theta}I)P_\theta$ is nilpotent of degree $d_\theta$. Lemma \ref{lemma:gregoir} shows that $D_\theta C_\theta=I$ and so
	\begin{align*}
		(\Phi-e^{-i\theta}I)P_\theta x_{t-1}&=\sum_{k=1}^{d_\theta-1}(\Phi-e^{-i\theta}I)^{k}P_\theta D_\theta(C_\theta B)^{k}[(R_\theta v_\leftrightarrow)_t+(C_\theta\varepsilon)_t]\\
		&=D_\theta\sum_{k=2}^{d_\theta}(\Phi-e^{-i\theta}I)^{k-1}P_\theta (C_\theta B)^{k-1}[(R_\theta v_\leftrightarrow)_t+(C_\theta\varepsilon)_t]
	\end{align*}
	for each $t\in\mathbb Z$. It therefore follows from \eqref{eq:Ptheta} and the definition of $D_\theta$ that
	\begin{align*}
		(\Phi-e^{-i\theta}I)P_\theta x_{t-1}&=D_\theta(P_\theta x_t-P_\theta [(R_\theta v_\leftrightarrow)_t+(C_\theta\varepsilon)_t])\\
		&=P_\theta x_t-P_\theta [(R_\theta v_\leftrightarrow)_t+(C_\theta\varepsilon)_t]\\
		&\quad\quad-e^{-i\theta}(P_\theta x_{t-1}-P_\theta[(R_\theta v_\leftrightarrow)_{t-1}+(C_\theta\varepsilon)_{t-1}])\quad\text{for each }t\in\mathbb Z.
	\end{align*}
	Adding $e^{-i\theta}P_\theta x_{t-1}$ to both sides of the last equation gives
	\begin{align*}
		\Phi P_\theta x_{t-1}&=P_\theta x_t-P_\theta[(R_\theta v_\leftrightarrow)_t+(C_\theta\varepsilon)_t]+e^{-i\theta}P_\theta[(R_\theta v_\leftrightarrow)_{t-1}+(C_\theta\varepsilon)_{t-1}]\\&=P_\theta x_t-P_\theta D_\theta[(R_\theta v_\leftrightarrow)_{t}+(C_\theta\varepsilon)_{t}]\quad\text{for each }t\in\mathbb Z.
	\end{align*}
	Lemma \ref{lemma:gregoir} shows that $D_\theta R_\theta=\bigzero$ and $D_\theta C_\theta=I$, so we have $P_\theta x_t=\Phi P_\theta x_{t-1}+P_\theta\varepsilon_t$ for each $t\in\mathbb Z$. This completes our demonstration that if \ref{en:statement2} is true then \ref{en:statement1} is true. We conclude that \ref{en:statement1} and \ref{en:statement2} are equivalent.
	
	It remains only to establish that if \ref{en:statement1} is true then \ref{en:statement4} is true. Suppose that \ref{en:statement1} is true and that $x$ is subexponential. We aim to show that $v_\rightarrow=0$ and $v_\leftarrow=0$. Note firstly that
	\begin{equation*}
		\lVert v_\rightarrow\rVert=\lim_{n\to\infty}\lVert\Phi^nP_\rightarrow x_{-n}\rVert\leq\limsup_{n\to\infty}\sum_{k=n}^\infty\lVert\Phi^kP_\rightarrow x_{-k}\rVert=\limsup_{n\to\infty}\sum_{k=n}^\infty\lVert(\Phi P_\rightarrow)^k x_{-k}\rVert,
	\end{equation*}
	where the final equality holds because $P_\rightarrow$ is idempotent and commutes with $\Phi$. Lemma \ref{lemma:spectralprojection} shows that $\Phi P_{\rightarrow}$ has all eigenvalues inside the unit circle. Therefore, since $x$ is subexponential, Gelfand's formula shows that the final limit superior is zero. Therefore $v_\rightarrow=0$. Note secondly that
	\begin{equation*}
		\lVert v_\leftarrow\rVert=\lim_{n\to\infty}\lVert\Phi^{-n}P_\leftarrow x_{n}\rVert\leq\limsup_{n\to\infty}\sum_{k=n}^\infty\lVert\Phi^{-k}P_\leftarrow x_{k}\rVert=\limsup_{n\to\infty}\sum_{k=n}^\infty\lVert(\Phi^\mathrm{D} P_\leftarrow)^k x_{k}\rVert,
	\end{equation*}
	where the final equality holds because $P_\leftarrow$ is idempotent and commutes with $\Phi^\mathrm{D}$. Lemma \ref{lemma:Drazin} shows that $\Phi^\mathrm{D} P_{\leftarrow}$ has all eigenvalues inside the unit circle. Therefore, since $x$ is subexponential, Gelfand's formula shows that the final limit superior is zero. Therefore $v_\leftarrow=0$. This completes our demonstration that \ref{en:statement1} implies \ref{en:statement4}.
\end{proof}

Theorem \ref{theorem:multivariate} establishes explicit formul{\ae} for the six flows in \eqref{eq:figure}. The six flows are given for each $t\in\mathbb Z$ by:
\begin{align}
	\Phi^t\lim_{n\to\infty}\Phi^n P_\rightarrow x_{-n}&\,\,\,\,\text{(predetermined forward }x\text{-flow)},\label{eq:predetforward}\\
	\sum_{k=0}^\infty\Phi^k P_{\dotrightarrow}\varepsilon_{t-k}&\,\,\,\,\text{(forward }\varepsilon\text{-flow)},\label{eq:forward}\\
	\Phi^t\lim_{n\to\infty}\Phi^{-n}P_\leftarrow x_n&\,\,\,\,\text{(predetermined backward }x\text{-flow)},\label{eq:predetbackward}\\
	-\sum_{k=1}^\infty\Phi^{-k}P_\leftarrow\varepsilon_{t+k}&\,\,\,\,\text{(backward }\varepsilon\text{-flow)},\label{eq:backward}\\
	\sum_{\theta\in\Theta}\sum_{k=1}^{d_\theta}(\Phi-e^{-i\theta}I)^{k-1}P_\theta((C_\theta B)^{k-1}R_\theta x_0)_t&\,\,\,\,\text{(predetermined outward }x\text{-flow)},\label{eq:predetoutward}\\
	\sum_{\theta\in\Theta}\sum_{k=1}^{d_\theta}(\Phi-e^{-i\theta}I)^{k-1}P_\theta((C_\theta B)^{k-1}C_\theta\varepsilon)_t&\,\,\,\,\text{(outward }\varepsilon\text{-flow)}.\label{eq:outward}
\end{align}
Every solution $x$ is the sum of these six flows. We emphasize that for $k\in\mathbb N$ the notation $\Phi^{-k}$ refers to the $k$th power of the Drazin inverse of $\Phi$. We will elaborate upon our terminology for the six flows in Section \ref{sec:arrow}.

The formul{\ae} in \eqref{eq:predetforward}--\eqref{eq:backward} involve only real matrices and vectors. If $\Theta$ includes some frequency $\theta\in(0,\pi)$ then the corresponding summands in \eqref{eq:predetoutward} and \eqref{eq:outward} are complex, as are those for the conjugate frequency $-\theta\in\Theta$. The sums evaluate to real vectors because the summands for conjugate frequencies are themselves conjugate. Indeed, the vectors
\begin{equation*}
	(\Phi-e^{-i\theta}I)^{k-1}P_\theta((C_\theta B)^{k-1}R_\theta x_0)_t\quad\text{and}\quad(\Phi-e^{i\theta}I)^{k-1}P_{-\theta}((C_{-\theta} B)^{k-1}R_{-\theta} x_0)_t
\end{equation*}
are complex conjugates for each $k\in\{1,\dots,d_\theta\}$ and each $t\in\mathbb Z$, and the same is true of the vectors
\begin{equation*}
	(\Phi-e^{-i\theta}I)^{k-1}P_\theta((C_\theta B)^{k-1}C_\theta\varepsilon)_t\quad\text{and}\quad(\Phi-e^{i\theta}I)^{k-1}P_{-\theta}((C_{-\theta} B)^{k-1}C_{-\theta}\varepsilon)_t.
\end{equation*}
The summation of complex conjugate vectors produces a real vector. Note that $d_{-\theta}=d_\theta$.

It is possible to re-express the formul{\ae} in \eqref{eq:predetoutward} and \eqref{eq:outward} using only real matrices and vectors. This can be done by using Euler's formula $e^{i\theta}=\cos\theta+i\sin\theta$ and the real versions of the frequency-specific difference and cumulation operators discussed in \citet[pp.~437--41]{Gregoir1999a}. The effect is to replace the sums of conjugate pairs in \eqref{eq:predetoutward} and \eqref{eq:outward} with real trigonometric expressions. The resulting formul{\ae} are in general much more complicated than those in \eqref{eq:predetoutward} and \eqref{eq:outward} and are omitted. For the particular case where a pair of conjugate eigenvalues on the unit circle has common index one we provide real trigonometric expressions for the corresponding outward flows in Remark \ref{rem:conjugate}.

\section{Remarks on Theorem \ref{theorem:multivariate} and related literature}\label{sec:remarks}

\begin{remark}\label{rem:nilpotent}
	Theorem \ref{theorem:multivariate} establishes a one-to-one correspondence between $\mathbb V_\rightarrow\times\mathbb V_\leftarrow\times\mathbb V_\leftrightarrow$ and the set of all solutions to \eqref{eq:VAR1}. Let $f_{\Phi,\varepsilon}$ be the map from $\mathbb V_\rightarrow\times\mathbb V_\leftarrow\times\mathbb V_\leftrightarrow$ into the set of all sequences in $\mathbb R^N$ that is defined by
	\begin{align*}
		f_{\Phi,\varepsilon}(v_\rightarrow,v_\leftarrow,v_\leftrightarrow)_t&=\Phi^tv_\rightarrow+\sum_{k=0}^\infty\Phi^k P_{\dotrightarrow}\varepsilon_{t-k}+\Phi^tv_\leftarrow-\sum_{k=1}^\infty\Phi^{-k}P_\leftarrow\varepsilon_{t+k}\\
		&\quad\quad+\sum_{\theta\in\Theta}\sum_{k=1}^{d_\theta}(\Phi-e^{-i\theta}I)^{k-1}P_\theta(C_\theta B)^{k-1}[(R_\theta v_\leftrightarrow)_t+(C_\theta\varepsilon)_t]
	\end{align*}
	for each $t\in\mathbb Z$. Thus $f_{\Phi,\varepsilon}$ sends each $(v_\rightarrow,v_\leftarrow,v_\leftrightarrow)\in\mathbb V_\rightarrow\times\mathbb V_\leftarrow\times\mathbb V_\leftrightarrow$ to the sum of the sequences defined by the right-hand sides of equations \eqref{eq:mainthm1}--\eqref{eq:mainthm3}. By construction, for each $(v_\rightarrow,v_\leftarrow,v_\leftrightarrow)\in\mathbb V_\rightarrow\times\mathbb V_\leftarrow\times\mathbb V_\leftrightarrow$, statement \ref{en:statement2} is true for the sequence $x=f_{\Phi,\varepsilon}(v_\rightarrow,v_\leftarrow,v_\leftrightarrow)$. Theorem \ref{theorem:multivariate} shows that \ref{en:statement2} implies \ref{en:statement1}. Thus $x$ must be a solution. This shows that the range of $f_{\Phi,\varepsilon}$ is a subset of the set of all solutions. Now let $x$ be any solution. Statement \ref{en:statement1} is true for this sequence $x$. Theorem \ref{theorem:multivariate} shows that \ref{en:statement1} implies \ref{en:statement2}. Thus $x=f_{\Phi,\varepsilon}(v_\leftarrow,v_\rightarrow,v_\leftrightarrow)$ for some $(v_\rightarrow,v_\leftarrow,v_\leftrightarrow)\in\mathbb V_\rightarrow\times\mathbb V_\leftarrow\times\mathbb V_\leftrightarrow$. This shows that the set of all solutions is a subset of the range of $f_{\Phi,\varepsilon}$. We conclude that the set of all solutions is equal to the range of $f_{\Phi,\varepsilon}$. Moreover, assertion \ref{en:statement3} in Theorem \ref{theorem:multivariate} shows that $f_{\Phi,\varepsilon}$ is injective. Thus $f_{\Phi,\varepsilon}$ defines a one-to-one correspondence between $\mathbb V_\rightarrow\times\mathbb V_\leftarrow\times\mathbb V_\leftrightarrow$ and the set of all solutions.
	
	The identity $P_\bullet+P_\rightarrow+P_\leftarrow+P_\leftrightarrow=I$ shows that the vector space $\mathbb V_\rightarrow\times\mathbb V_\leftarrow\times\mathbb V_\leftrightarrow$ has dimension $N-\dim(\mathbb V_\bullet)$. Consequently there is exactly one solution to \eqref{eq:VAR1} if and only if $\dim(\mathbb V_\bullet)=N$; that is, if and only if $\Phi$ is nilpotent. In this case $P_\bullet=I$, $P_\rightarrow=P_\leftarrow=P_\leftrightarrow=\bigzero$, and the unique solution is given by
	\begin{equation*}
		x_t=f_{\Phi,\varepsilon}(0,0,0)_t=\sum_{k=0}^{N-1}\Phi^k\varepsilon_{t-k}\quad\text{for each }t\in\mathbb Z.
	\end{equation*}
	Otherwise there are infinitely many solutions.
	
	It is possible to regard $\mathbb V_\rightarrow\times\mathbb V_\leftarrow\times\mathbb V_\leftrightarrow$ as a space of \emph{initial conditions}. Each choice of $(v_\rightarrow,v_\leftarrow,v_\leftrightarrow)$ corresponds to a solution $x=f_{\Phi,\varepsilon}(v_\rightarrow,v_\leftarrow,v_\leftrightarrow)$ for which the projected sequences $P_\rightarrow x$, $P_\leftarrow x$ and $P_\leftrightarrow x$ are respectively constrained by the choices of $v_\rightarrow$, $v_\leftarrow$ and $v_\leftrightarrow$. The choice of $v_\rightarrow$ constrains the behavior of $P_\rightarrow x$ in the arbitrarily distant past. The choice of $v_\leftarrow$ constrains the behavior of $P_\leftarrow x$ in the arbitrarily distant future. The choice of $v_\leftrightarrow$ constrains the behavior of $P_\leftrightarrow x$ at time zero. Therefore, if we imagine $P_\rightarrow x$, $P_\leftarrow x$ and $P_\leftrightarrow x$ to be respectively flowing forward, backward and outward in time, then the choices of $v_\rightarrow$, $v_\leftarrow$, $v_\leftrightarrow$ constrain the \emph{initial} behavior of the respective sequences. We elaborate on these ideas in Section \ref{sec:arrow}, drawing a connection to the description of the arrow of time given in \cite{Eddington1929}. If a more conventional perspective on time is preferred then we might instead say that the choices of $v_\rightarrow$ and $v_\leftarrow$ are \emph{boundary conditions at infinity} and reserve the term \emph{initial condition} for the choice of $v_\leftrightarrow$.
\end{remark}

\begin{remark}\label{rem:subexponentialepsflows}
	For each solution $x$, the forward, backward and outward $\varepsilon$-flows defined in \eqref{eq:forward}, \eqref{eq:backward} and \eqref{eq:outward}, and the predetermined outward $x$-flow defined in \eqref{eq:predetoutward}, are subexponential sequences. Consider the forward $\varepsilon$-flow. Note firstly that all eigenvalues of $\Phi P_{\dotrightarrow}$ are inside the unit circle. Let $\rho(\cdot)$ and $\lvvvert\cdot\rvvvert$ respectively be the spectral radius and spectral norm of a square matrix.\footnote{The spectral norm of a real $N\times N$ matrix $M$ is defined by $\lvvvert M\rvvvert=\sup_{v\in\mathbb R^N\setminus\{0\}}\lVert Mv\rVert/\lVert v\rVert$. See \citet[pp.~493--4]{BanerjeeRoy2014}} Choose $r\in(\rho(\Phi P_{\dotrightarrow}),1)$ and note that $r^{\lvert t\rvert}\leq r^{\lvert t-k\rvert}\cdot r^{-k}$ for each $t\in\mathbb Z$ and each nonnegative integer $k$. It follows that
	\begin{align*}
		\sum_{t\in\mathbb Z}r^{\lvert t\rvert}\left\Vert\sum_{k=0}^\infty\Phi^kP_{\dotrightarrow}\varepsilon_{t-k}\right\Vert&=\sum_{t\in\mathbb Z}r^{\lvert t\rvert}\left\Vert\sum_{k=0}^\infty(\Phi P_{\dotrightarrow})^k\varepsilon_{t-k}\right\Vert\\&\leq\sum_{k=0}^\infty r^{-k}\lvvvert (\Phi P_{\dotrightarrow})^k\rvvvert\sum_{t\in\mathbb Z}r^{\lvert t-k\rvert}\left\Vert\varepsilon_{t-k}\right\Vert.
	\end{align*}
	We have $\sum_{t\in\mathbb Z}r^{\lvert t-k\rvert}\left\Vert\varepsilon_{t-k}\right\Vert=\sum_{t\in\mathbb Z}r^{\lvert t\rvert}\left\Vert\varepsilon_{t}\right\Vert<\infty$ for each $k\in\mathbb N$ because $\varepsilon$ is subexponential, and we have $\sum_{k=0}^\infty r^{-k}\lvvvert(\Phi P_{\dotrightarrow})^k\rvvvert<\infty$ by Gelfand's formula because $r>\rho(\Phi P_{\dotrightarrow})$. Therefore the forward $\varepsilon$-flow is subexponential.  A similar argument based on the fact that all eigenvalues of $\Phi^\mathrm{D}P_\leftarrow$ are inside the unit circle shows that the backward $\varepsilon$-flow is subexponential. The outward $\varepsilon$-flow is subexponential because $C_\theta$ maps subexponential sequences to subexponential sequences: given any subexponential $y\in\mathbb S$ and any $r\in(0,1)$, we have
	\begin{align*}
		\sum_{t\in\mathbb Z}r^{\lvert t\rvert}\lVert(C_\theta y)_t\rVert&=\sum_{t\in\mathbb N}r^{t}\left\Vert\sum_{s=1}^{t}e^{-i\theta(t-s)}y_s\right\Vert+\sum_{t\in\mathbb N}r^{t}\left\Vert\sum_{s=0}^{t-1}e^{-i\theta(s-t)}y_{-s}\right\Vert\\
		&\leq\sum_{t\in\mathbb N}r^{t}\sum_{s=-t+1}^t\lVert y_s\rVert\leq\left(\sum_{t\in\mathbb N}r^{t/2}\right)\left(\sum_{s\in\mathbb Z}r^{\lvert s\rvert/2}\lVert y_s\rVert\right)<\infty.
	\end{align*}
	The predetermined outward $x$-flow is subexponential for the same reason. In fact, our subsequent discussion in Remark \ref{rem:predetoutward} shows that the predetermined outward $x$-flow is uniformly bounded in norm by a polynomial function of time.
\end{remark}

\begin{remark}\label{rem:predetforwardbackward}
	The sum of two subexponential sequences is itself subexponential. Every solution $x$ is the sum of the six flows in \eqref{eq:predetforward}--\eqref{eq:predetbackward}, and Remark \ref{rem:subexponentialepsflows} shows that four of these flows are subexponential. Only the predetermined forward and backward $x$-flows need not be subexponential. Thus the particular solution $x=f_{\Phi,\varepsilon}(0,0,v_\leftrightarrow)$ is subexponential for every $v_\leftrightarrow\in\mathbb V_\leftrightarrow$. Consequently assertion \ref{en:statement4} in Theorem \ref{theorem:multivariate} may be replaced with the following stronger assertion.
	\begin{enumerate}[label={\upshape(\roman*${'}$)},align=left]
		\setcounter{enumi}{\value{counter:items}-1}
		\item $x$ is subexponential if and only if $v_\rightarrow=0$ and $v_\leftarrow=0$.\label{en:statement4'}
	\end{enumerate}
	We further observe that, for each particular solution $x=f_{\Phi,\varepsilon}(v_\rightarrow,v_\leftarrow,v_\leftrightarrow)$, the predetermined forward $x$-flow is subexponential if and only if $v_\rightarrow=0$, and the predetermined backward $x$-flow is subexponential if and only if $v_\leftarrow=0$. The former observation follows from the fact that if $v_\rightarrow\neq0$ then $f_{\Phi,\varepsilon}(v_\rightarrow,0,0)$ is not subexponential and is the sum of the predetermined forward $x$-flow and the three subexponential $\varepsilon$-flows, and the latter observation follows from the fact that if $v_\leftarrow\neq0$ then $f_{\Phi,\varepsilon}(0,v_\leftarrow,0)$ is not subexponential and is the sum of the predetermined backward $x$-flow and the three subexponential $\varepsilon$-flows.
	
	Assertion \ref{en:statement4'} shows that the map $f_{\Phi,\varepsilon}(0,0,\cdot)$ defines a one-to-one correspondence between $\mathbb V_\leftrightarrow$ and the set of all subexponential solutions. If $\Phi$ has no eigenvalues on the unit circle then $\mathbb V_\leftrightarrow=\{0\}$ and there is exactly one subexponential solution. This algebraic fact, relying in no way on probabilistic concepts for its justification, underlies the well-known result stating that an autoregressive law of motion with no eigenvalues on the unit circle and with stationary innovations with finite expected norm admits a unique stationary solution.
\end{remark}

\begin{remark}\label{rem:predetoutward}
	Suppose that $\Phi$ has a unit eigenvalue and has no other eigenvalues on the unit circle, so that $\Theta=\{0\}$. In this case the formula for the predetermined outward $x$-flow in \eqref{eq:predetoutward} may be simplified. For the vector $x_0\in\mathbb R^N$ the definition shows $R_0x_0$ is a constant sequence with $(R_0x_0)_t=x_0$ for all $t\in\mathbb Z$. Successive applications of the operator $C_0B$ show that
	\begin{equation*}
		((C_0B)^{k-1}R_0x_0)_t={t\choose k-1}x_0\quad\text{for each }t\in\mathbb Z\text{ and each }k\in\mathbb N,
	\end{equation*}
	where ${p\choose q}=p\cdots(p-q+1)/q!$ denotes the generalized binomial coefficient. Therefore
	\begin{equation*}
		\sum_{k=1}^{d_0}(\Phi-I)^{k-1}P_0((C_0 B)^{k-1}R_0 x_0)_t=\sum_{k=1}^{d_0}{t\choose k-1}(\Phi-I)^{k-1}P_0x_0\quad\text{for each }t\in\mathbb Z,
	\end{equation*}
	which represents a substantial simplification to \eqref{eq:predetoutward} and shows that if $\Theta=\{0\}$ then the predetermined outward $x$-flow is a polynomial in time $t\in\mathbb Z$ with vector coefficients, with the degree of this polynomial no greater than $d_0-1$.
	
	If eigenvalues are permitted anywhere on the unit circle then the predetermined outward $x$-flow need not be a polynomial in $t$, but must nevertheless be uniformly bounded in norm by a polynomial in $t$. This may be seen by noting that $\lVert (R_\theta x_0)_t\rVert\leq\lVert x_0\rVert$ for each $t\in\mathbb Z$ and each $\theta\in\Theta$, and then applying arguments similar to those above.
\end{remark}

\begin{remark}\label{rem:HD}
	Let $\varepsilon$ be stationary random sequence in $\mathbb R^N$ with finite expected norm, and assume that $\Phi$ has no eigenvalues on the unit circle, so that $\Theta$ is empty. It is explained in \citet[pp.~9--12]{HannanDeistler1988} that in this case there exists a unique stationary solution $\tilde{x}$, and the set of all solutions is obtained by adding to $\tilde{x}$ any solution $y$ to the homogeneous difference equation $y_t=\Phi y_{t-1}$. See also Lemma 2 in \cite{Deistler1975} and Propositions 4.22 and 4.23 in \cite{DeistlerScherrer2022}. This description of the set of all solutions can be understood in terms of the six flows in \eqref{eq:predetforward}--\eqref{eq:outward} in the following way. The requirement that $\Phi$ has no eigenvalues on the unit circle eliminates the two outward flows because $\Theta$ is empty. The unique stationary solution $\tilde{x}$ is the sum of the forward and backward $\varepsilon$-flows. The predetermined forward and backward $x$-flows are both solutions to the homogeneous difference equation $y_t=\Phi y_{t-1}$ due to properties of the Drazin inverse and thus their sum is also a solution.
\end{remark}

\begin{remark}\label{rem:BHansen}
	Let $\varepsilon$ be stationary and ergodic with finite expected norm, and assume that all eigenvalues of $\Phi$ are inside the unit circle. Theorem 15.6 in \citet[p.~529]{Hansen2022}, a recent textbook on econometrics aimed at graduate students, asserts that in this case any solution to \eqref{eq:VAR1} is stationary and ergodic. The assertion is correct under the additional requirement -- prohibitively restrictive for typical applications -- that $\Phi$ is nilpotent. Absent nilpotency, one can only pin down a stationary and ergodic solution to \eqref{eq:VAR1} by imposing a suitable initial condition. The general form of all solutions is
	\begin{equation}
		x_t=\Phi^tv_\rightarrow+\sum_{k=0}^\infty\Phi^k\varepsilon_{t-k}\quad\text{for each }t\in\mathbb Z,\label{eq:HansenDrazin}
	\end{equation}
	where it should be understood that for negative $t\in\mathbb Z$ the notation $\Phi^t$ refers to $(\Phi^\mathrm{D})^{-t}$, and where $v_\rightarrow$ can be any real $N\times 1$ vector in the linear span of the generalized eigenvectors of $\Phi$ associated with nonzero eigenvalues. If $\Phi$ is nilpotent then all eigenvalues of $\Phi$ are zero and $\mathbb V_\rightarrow=\{0\}$. Thus we must set $v_\rightarrow=0$ in \eqref{eq:HansenDrazin}, thereby obtaining the unique solution to \eqref{eq:VAR1}. This solution is stationary and ergodic. If $\Phi$ is not nilpotent then $\Phi$ has a nonzero eigenvalue and thus there are infinitely many possible choices of $v_\rightarrow$ in \eqref{eq:HansenDrazin}, each yielding a different solution. One must choose $v_\rightarrow=0$ to obtain a stationary and ergodic solution. Choosing $v_\rightarrow=0$ is equivalent to imposing the initial condition $\lim_{n\to\infty}\Phi^nx_{-n}=0$.
	
	The source of the problem in Theorem 15.6 in \cite{Hansen2022} can be found in Theorem 14.21 on p.~478 therein, and in the preceding discussion. Here backward recursion on $x_t=\Phi x_{t-1}+\varepsilon_t$ is used to deduce that
	\begin{equation}\label{eq:BHansen1}
		x_t=\Phi^tx_0+\sum_{k=0}^{t-1}\Phi^k\varepsilon_{t-k}\quad\text{for each }t\in\mathbb N.
	\end{equation}
	This is the essence of Eq.~14.26 in \citet[p.~478]{Hansen2022}. It is then argued, by appealing to Theorem 14.3 in \citet[p.~461]{Hansen2022}, that if this recursion is continued into \emph{the infinite past} one obtains $x_t=\sum_{k=0}^\infty\Phi^k\varepsilon_{t-k}$. This is not correct in general. Theorem 14.3 correctly asserts that for each $t\in\mathbb Z$ the series $\sum_{k=0}^\infty\Phi^k\varepsilon_{t-k}$ converges (with probability one). To apply this result we may use backward recursion on $x_t=\Phi x_{t-1}+\varepsilon_t$ to write
	\begin{equation}\label{eq:BHansen2}
		x_t=\Phi^{n+1}x_{t-n-1}+\sum_{k=0}^{n}\Phi^k\varepsilon_{t-k}\quad\text{for each }t\in\mathbb Z\text{ and each }n\in\mathbb N\cup\{0\}.
	\end{equation}
	Note that \eqref{eq:BHansen1} may be recovered from \eqref{eq:BHansen2} by setting $n=t-1$. Theorem 14.3 establishes that $\sum_{k=0}^\infty\Phi^k\varepsilon_{t-k}$ is the limit as $n\to\infty$ of the second term on the right-hand side of \eqref{eq:BHansen2}. However, despite the requirement that $\Phi$ has all eigenvalues inside the unit circle, it is not in general the case that the first term on the right-hand side of \eqref{eq:BHansen2} converges to zero as $n\to\infty$. Indeed, if $\Phi$ is not nilpotent then for all but one of the infinitely many solutions to \eqref{eq:VAR1} we have $\lim_{n\to\infty}\Phi^{n+1}x_{t-n-1}\neq0$ for each $t\in\mathbb Z$, and the limit grows exponentially in norm as $t\to-\infty$.
	
	The same issue arises in \citet{Lutkepohl2005}, another econometrics textbook aimed at graduate students. The substantive content of \eqref{eq:BHansen2} appears on p.~14 where, in our notation, it is asserted that $\Phi^{n+1}$ converges to zero rapidly as $n\to\infty$ and thus one may \emph{ignore the term $\Phi^{n+1}x_{t-n-1}$ in the limit}. As we have seen, the rapid convergence of $\Phi^{n+1}$ to zero does not imply convergence of $\Phi^{n+1}x_{t-n-1}$ to zero. Ignoring this term leads to the assertion in Proposition 2.1 in \citet[p.~25]{Lutkepohl2005} that an autoregressive process is stationary if $\Phi$ has all eigenvalues inside the unit circle. This assertion is only correct on the understanding that solutions for which $\Phi^{n+1}x_{t-n-1}$ does not converge to zero as $n\to\infty$ are excluded from consideration.
	
	Our flow decomposition may illuminate the issue just discussed. When all eigenvalues of $\Phi$ are inside the unit circle the backward and outward flows in \eqref{eq:predetbackward}--\eqref{eq:outward} are zero. This leaves us with the predetermined forward $x$-flow and the forward $\varepsilon$-flow. In both \citet{Lutkepohl2005} and \citet{Hansen2022} the predetermined forward $x$-flow is excluded. Consequently the forward $\varepsilon$-flow, which is stationary and ergodic when $\varepsilon$ is stationary and ergodic, is taken to be the unique representation of $x$. See Remark \ref{rem:Hansen} for another case where the predetermined forward $x$-flow has been implicitly excluded.
	
	While it is not true that every solution to \eqref{eq:VAR1} is stationary when $\Phi$ has all eigenvalues inside the unit circle and $\varepsilon$ is stationary with finite expected norm, there is a sense in which every solution is \emph{asymptotically} stationary. The reason is that the predetermined forward $x$-flow converges exponentially to zero as time progresses. If one adopts the conventional statistical perspective in which $x$ is observed at times $t\in\{0,1,\dots,n\}$ and justifies the use of a statistical procedure via an asymptotic approximation as $n\to\infty$, the presence of the predetermined $x$-flow is typically irrelevant to this approximation. On the other hand, if one were to adopt a contrary perspective in which the process is observed at times $t\in\{-n,-n+1,\dots,0\}$, then the corresponding asymptotic approximation as $n\to\infty$ would typically be dominated by the predetermined forward $x$-flow, which grows exponentially as time regresses.
\end{remark}

\begin{remark}\label{rem:Potscher}
	Consider the univariate case in which $\Phi$ is a real number $\phi$, and assume that $\lvert\phi\rvert>1$ and that the innovations are independent and identically distributed with zero mean and finite expected norm. In this case the unique stationary solution $\tilde{x}$ to \eqref{eq:VAR1} is the anti-causal linear process $\tilde{x}_t=-\sum_{k=1}^\infty\phi^{-k}\varepsilon_{t+k}$; see, for instance, \citet[p.~18]{Hannan1970}, \citet[p.~134]{Priestley1981}, \citet[p.~81]{BrockwellDavis1991}, \citet[p.~377]{Hayashi2000} or \citet[pp.~93--4]{Rosenblatt2000}. Other prominent textbooks are less clear on this matter, sometimes asserting that if $\lvert\phi\rvert>1$ then stationary solutions do not exist, or that if $\lvert\phi\rvert>1$ then every solution is explosive, leaving tacit whatever assumptions may justify these claims. Examples include \citet[p.~270]{DavidsonMacKinnon2004}, \citet[pp.~39, 402]{Tsay2010} and \citet[p.~481]{Hansen2022}. On p.~41 in \cite{Hamilton1994} it is advised that one solve an autoregressive law of motion \emph{backward} if $\lvert\phi\rvert<1$ or \emph{forward} if $\lvert\phi\rvert>1$. Following this advice, which is attributed to \citet{Sargent1987}, does indeed identify the unique stationary solution $\tilde{x}_t=-\sum_{k=1}^\infty\phi^{-k}\varepsilon_{t+k}$ for the case $\lvert\phi\rvert>1$. However it is then stated on p.~53 in \cite{Hamilton1994} that covariance-stationary solutions to an autoregressive law of motion with $\lvert\phi\rvert\geq1$ do not exist, leaving tacit the requirement that the solution space be confined to causal linear processes. See also \cite{Potscher1996}.
	
	While it is not true that every solution is explosive in the univariate case with $\lvert\phi\rvert>1$, there is an important intuitive element to this statement which our flow decomposition may help to illuminate. When $\lvert\phi\rvert>1$ the forward and outward flows in \eqref{eq:predetforward}, \eqref{eq:forward}, \eqref{eq:predetoutward} and \eqref{eq:outward} are zero, so each solution $x$ is the sum of the predetermined backward $x$-flow in \eqref{eq:predetbackward} and the backward $\varepsilon$-flow in \eqref{eq:backward}. For exactly one solution $x$, out of infinitely many solutions, the predetermined backward $x$-flow is zero and $x$ is equal to the backward $\varepsilon$-flow, a stationary anti-causal linear process. For each of the infinitely many other solutions the predetermined backward $x$-flow is nonzero and grows exponentially as time progresses. Thus every solution grows exponentially as time progresses except for one particular solution, the unique stationary solution. Moreover, as discussed in \cite{GourierouxZakoian2017}, even the unique stationary solution may be viewed as exhibiting a form of local explosivity.
\end{remark}

\begin{remark}
	The outward $\varepsilon$-flow in \eqref{eq:outward} has been a central object of study in the econometric literature on unit roots, where it is commonly called a stochastic trend. The dominant focus of the literature has been the case where $\Theta=\{0\}$ and $d_0=1$. In this case, referred to as the $I(1)$ case, the outward $\varepsilon$-flow is simply $P_0 C_0\varepsilon$: a projection of cumulated innovations on the eigenspace of $\Phi$ associated with its unit eigenvalue. Significant attention has also been devoted to the case where $\Theta=\{0\}$ and $d_0=2$. In this case, referred to as the $I(2)$ case, the outward $\varepsilon$-flow is $P_0 C_0\varepsilon+(\Phi-I)P_0C_0BC_0\varepsilon$. Expressions provided in prior literature for the outward $\varepsilon$-flow in the $I(2)$ case have been much more complicated in form. See, for instance, \citet[pp.~125--6]{Johansen2008} and \citet[p.~786]{BeareSeo2020}. The main reason for this is that prior literature has relied on the use of orthogonal projections or complements to study the structure of the outward $\varepsilon$-flow. It is much cleaner to work with the spectral projections, which are typically not orthogonal projections. For cases where $\Theta=\{0\}$ and $d_0\geq3$, \citet[p.~1184]{FranchiParuolo2019} resorts to a recursive characterization of the outward $\varepsilon$-flow wherein one starts with an expression for the final term in the sum over $k$ in \eqref{eq:outward} and then applies an iterative procedure to derive expressions for each of the preceding terms. We see from \eqref{eq:outward} that it is simpler to start with the first term in the sum, $P_0C_0\varepsilon$, and then repeatedly apply $(\Phi-I)C_0B$ to obtain all following terms. In \citet[p.~435]{HowlettBeareFranchiBolandAvrachenkov2024} an expression for the outward $\varepsilon$-flow in cases where $\Theta=\{0\}$ and $d_0\in\mathbb N\cup\{\infty\}$ is provided in terms of the spectral projection associated with the unit eigenvalue, similar to what is done here. There the setting is a Banach space with possibly infinite dimension, and the case $d_0=\infty$ occurs when there is an infinite-length Jordan chain of generalized eigenvectors associated with the unit eigenvalue. Confining attention to the finite-dimensional setting $\mathbb R^N$, the expression for the outward $\varepsilon$-flow given in \eqref{eq:outward} extends the one given in \cite{HowlettBeareFranchiBolandAvrachenkov2024} by allowing $\Theta$ to be unrestricted and by not confining time to the nonnegative integers.
\end{remark}

\begin{remark}\label{rem:conjugate}
	The econometric literature on seasonal unit roots has focused attention on cases where $\Theta=\{\theta,-\theta\}$ for some nonzero $\theta\in(-\pi,\pi)$ and where $d_\theta=d_{-\theta}=1$; that is, on cases where $\Phi$ has exactly two eigenvalues on the unit circle and these form a conjugate pair with common index one. In such cases the formul{\ae} for the predetermined outward $x$-flow and the outward $\varepsilon$-flow in \eqref{eq:predetoutward} and \eqref{eq:outward} respectively simplify to $((P_\theta R_\theta +P_{-\theta}R_{-\theta}) x_0)_t$ and $((P_\theta C_\theta+P_{-\theta} C_{-\theta})\varepsilon)_t$ for each $t\in\mathbb Z$. We may rewrite the simplified expression for the predetermined outward $x$-flow in the real trigonometric form
	\begin{equation*}
		\big(\cos(\theta t)(P_\theta+P_{-\theta})+\sin(\theta t)[(-i)(P_\theta-P_{-\theta})]\big)x_0\quad\text{for each }t\in\mathbb Z
	\end{equation*}
	by applying the identities
	\begin{equation*}
		(R_\theta x_0)_t=(\cos(\theta t)-i\sin(\theta t))x_0\quad\text{and}\quad(R_{-\theta} x_0)_t=(\cos(\theta t)+i\sin(\theta t))x_0
	\end{equation*}
	obtained from Euler's formula $e^{i\theta}=\cos(\theta)+i\sin(\theta)$. Note that $P_\theta$ and $P_{-\theta}$ are complex conjugate matrices by part \ref{en:real} of Lemma \ref{lemma:spectralprojection}, and thus $P_\theta+P_{-\theta}$ and $(-i)(P_\theta-P_{-\theta})$ are real matrices. A similar application of Euler's formula shows that the simplified expression for the outward $\varepsilon$-flow may be rewritten in the real trigonometric form
	\begin{equation*}
		\sum_{s=1}^t\big(\cos(\theta(t-s))(P_\theta+P_{-\theta})+\sin(\theta(t-s))[(-i)(P_\theta-P_{-\theta})]\big)\varepsilon_s
	\end{equation*}
	for all positive $t\in\mathbb Z$, and in the real trigonometric form
	\begin{equation*}
		-\sum_{s=0}^{-t-1}\big(\cos(\theta(t+s))(P_\theta+P_{-\theta})+\sin(\theta(t+s))[(-i)(P_\theta-P_{-\theta})]\big)\varepsilon_{-s}
	\end{equation*}
	for all negative $t\in\mathbb Z$. At time $t=0$ the outward $\varepsilon$-flow is zero.
\end{remark}

\begin{remark}\label{rem:Hansen}
	In literature dealing with the leading case of interest where $\Theta=\{0\}$, $d_0=1$ and $\Phi$ has no eigenvalues outside the unit circle, the Granger-Johansen representation theorem is often said to decompose an autoregressive process with white noise innovations into the sum of three parts: a random walk, a stationary component, and an initial condition (and possibly a fourth part related to the inclusion of additional nonrandom terms in the autoregressive law of motion, which may be ignored for the present discussion). See, for instance, \citet[p.~23]{Hansen2005}. In our notation, Theorem 1 therein states that if $x$ satisfies $x_t=\Phi x_{t-1}+\varepsilon_t$ for each $t\in\mathbb N$ then
	\begin{equation}
		x_t=P_0(C_0\varepsilon)_t+\sum_{k=0}^\infty\Phi^kP_{\dotrightarrow}\varepsilon_{t-k}+P_0x_0\quad\text{for each }t\in\mathbb N\label{eq:Hansen}
	\end{equation}
	The innovations $\varepsilon_t$ are defined for all $t\in\mathbb Z$ while, following the usual practice in econometric literature, $x_t$ is defined only for nonnegative $t\in\mathbb Z$.
	
	The three terms on the right-hand side of \eqref{eq:Hansen} are, respectively, the aforementioned random walk, stationary component, and initial condition. They are respectively equal to the outward $\varepsilon$-flow in \eqref{eq:outward}, the forward $\varepsilon$-flow in \eqref{eq:forward}, and the predetermined outward $x$-flow in \eqref{eq:predetoutward}. The predetermined backward $x$-flow and backward $\varepsilon$-flow are absent because of the assumption that $\Phi$ has no eigenvalues outside the unit circle. What has happened to the predetermined forward $x$-flow? A close reading of the proof of Theorem 1 in \citet{Hansen2005} reveals that in the second paragraph the law of motion $P_{\dotrightarrow}x_t=\Phi P_{\dotrightarrow}x_{t-1}+P_{\dotrightarrow}\varepsilon_t$ is used to justify $P_{\dotrightarrow}x_t$ having the \emph{stationary representation} $P_{\dotrightarrow}x_t=\sum_{k=0}^\infty\Phi^kP_{\dotrightarrow}\varepsilon_{t-k}$. But the last equality is not true in general. After $t$ iterations of the law of motion for $P_{\dotrightarrow}x_t$ we have $P_{\dotrightarrow}x_t=\Phi^tP_{\dotrightarrow}x_0+\sum_{k=0}^{t-1}P_{\dotrightarrow}\varepsilon_{t-k}$. Further iterations are not possible unless we define $P_{\dotrightarrow}x_t$ for negative $t\in\mathbb Z$, and assume that the law of motion $P_{\dotrightarrow}x_t=\Phi P_{\dotrightarrow}x_{t-1}+P_{\dotrightarrow}\varepsilon_t$ is satisfied for nonpositive $t\in\mathbb Z$. If we do then we obtain \eqref{eq:mainthmprf00} for each $t\in\mathbb Z$ and each $n\in\mathbb N$, and deduce by arguing as in the proof of Theorem \ref{theorem:multivariate} that \eqref{eq:mainthm1} is satisfied with $v_\rightarrow=\lim_{n\to\infty}\Phi^nP_\rightarrow x_{-n}$. Thus \eqref{eq:Hansen} should be amended to read
	\begin{equation}
		x_t=P_0(C_0\varepsilon)_t+\sum_{k=0}^\infty\Phi^kP_{\dotrightarrow}\varepsilon_{t-k}+P_0x_0+\Phi^t\lim_{n\to\infty}\Phi^nP_\rightarrow x_{-n},\label{eq:Hansenfix}
	\end{equation}
	with the equality now holding for each $t\in\mathbb Z$, and with the understanding that for negative $t\in\mathbb Z$ the notation $\Phi^t$ refers to $(\Phi^\mathrm{D})^{-t}$. If we are not willing to define $P_{\dotrightarrow}x_t$ at negative times then there is no basis for deducing that $P_{\dotrightarrow}x_t=\sum_{k=0}^\infty\Phi^kP_{\dotrightarrow}\varepsilon_{t-k}$ for each $t\in\mathbb N$, and to obtain \eqref{eq:Hansen} we must assume that
	\begin{equation}
		P_{\dotrightarrow}x_0=\sum_{k=0}^\infty\Phi^kP_{\dotrightarrow}\varepsilon_{-k}.\label{eq:Hansenlem}
	\end{equation}
	In fact, \eqref{eq:Hansenlem} is Lemma 2 in \citet{Hansen2005}, which is neither proved nor explicitly used in the proof of Theorem 1 therein. It may be viewed as an assumption used implicitly in the proof of Theorem 1. Immediately following the statement of Lemma 2 in \citet{Hansen2005}, an alternative statement of the conclusion of Theorem 1 is given in which \eqref{eq:Hansen} is replaced with
	\begin{equation*}
		x_t=P_0(C_0\varepsilon)_t+\sum_{k=0}^\infty\Phi^kP_{\dotrightarrow}\varepsilon_{t-k}+x_0-\sum_{k=0}^\infty\Phi^kP_{\dotrightarrow}\varepsilon_{-k}\quad\text{for each }t\in\mathbb N,
	\end{equation*}
	as in \citet{Johansen1991}. This equality reduces to \eqref{eq:Hansen} if \eqref{eq:Hansenlem} is assumed to hold.
	
	The issues just raised are not unique to \citet{Hansen2005}. Most other literature on the Granger-Johansen representation theorem has omitted or obscured the predetermined forward $x$-flow; we have singled out \citet{Hansen2005} because the clarity of writing makes it possible to see where the flow is implicitly assumed to be zero. In the statement of the Granger-Johansen representation theorem in \citet[p.~49]{Johansen1995}, the phrase \emph{can be given initial distributions} is used to implicitly exclude the predetermined forward $x$-flow from consideration. Indeed, if one requires that $x_0$ satisfies \eqref{eq:Hansenlem} then the predetermined forward $x$-flow must be zero, as can be seen by setting $t=0$ in \eqref{eq:Hansenfix}.
\end{remark}

\begin{remark}
	Theorem 1 in \citet{Nielsen2010} extends the Granger-Johansen representation theorem to cases where $\Phi$ has a single eigenvalue $\lambda$ outside the unit circle. The representation provided for $x_t$ includes a stochastic exponential trend of the form $\sum_{s=1}^t\lambda^{t-s}\varepsilon_s$. This trend does not obviously resemble any of the six flows in the representation for $x_t$ we have provided in Theorem \ref{theorem:multivariate}. We may nevertheless deduce Nielsen's representation from ours in the following way. Use \eqref{eq:mainthm2} in Theorem \ref{theorem:multivariate} to write
	\begin{equation*}
		\Phi^tP_\leftarrow x_0=\Phi^t\lim_{n\to\infty}(\Phi^\mathrm{D})^nP_\leftarrow x_n-\Phi^t\sum_{k=1}^\infty(\Phi^\mathrm{D})^{k}P_\leftarrow\varepsilon_k\quad\text{for each }t\in\mathbb N.
	\end{equation*}
	Subtracting the last equation from \eqref{eq:mainthm2} gives
	\begin{align*}
		P_\leftarrow x_t&=\Phi^tP_\leftarrow x_0+\Phi^t\sum_{k=1}^\infty(\Phi^\mathrm{D})^{k}P_\leftarrow\varepsilon_k-\sum_{k=1}^\infty(\Phi^\mathrm{D})^{k}P_\leftarrow\varepsilon_{t+k}\\
		&=\Phi^tP_\leftarrow x_0+\Phi^t\sum_{k=1}^\infty(\Phi^\mathrm{D})^{k}P_\leftarrow\varepsilon_k-\Phi^t\sum_{k=1}^\infty(\Phi^\mathrm{D})^{t+k}P_\leftarrow\varepsilon_{t+k}\\
		&=\Phi^tP_\leftarrow x_0+\Phi^t\sum_{s=1}^t(\Phi^\mathrm{D})^{s}P_\leftarrow\varepsilon_{s}=\Phi^tP_\leftarrow x_0+\sum_{s=1}^t\Phi^{t-s}P_\leftarrow\varepsilon_{s}\quad\text{for each }t\in\mathbb N,
	\end{align*}
	where we have used \eqref{eq:Drazin2} and part \ref{en:restrictDrazin} of Lemma \ref{lemma:Drazin} to obtain the second and fourth equalities. The two terms on the right-hand side of the final equality correspond to the second and sixth terms in the representation for $x_t$ provided by Theorem 1 in \citet{Nielsen2010}. One may be viewed as a stochastic exponential trend depending on the innovations at times $1$ through $t$. It is tempting to view the other as a predetermined (i.e., determined by $P_\leftarrow x_0$) exponential trend. However, $P_\leftarrow x_0$ is itself dependent on the innovations at all positive times, as shown by setting $t=0$ in \eqref{eq:mainthm2}. In the backward space $\mathbb V_\leftarrow$ it is more natural to think of a sequence as being predetermined when it is determined in the arbitrarily distant future. We elaborate on this idea in the following section.
\end{remark}

\section{Measurability and the arrow of time}\label{sec:arrow}

The terminology we have assigned to our six flows in \eqref{eq:predetbackward}--\eqref{eq:outward} is motivated by, and may be formalized using, the concept of measurability. Let $(\Omega,\mathcal A)$ and $(R,\mathcal R)$ be measurable spaces. From $(R,\mathcal R)$ we construct a third measurable space $(S,\mathcal S)$ by taking $S$ to be the set of all sequences in $R$ indexed by $t\in\mathbb Z$, and by taking $\mathcal S$ to be the product sigma-algebra on $S$, i.e.\ the coarsest sigma-algebra under which $s\mapsto s_t$ is a measurable map from $S$ to $R$ for each $t\in\mathbb Z$.

We have in mind situations where $(\Omega,\mathcal A)$ is equipped with a probability measure and where $(R,\mathcal R)$ is the Euclidean space $\mathbb R^N$ together with its Borel sigma-algebra. In such cases a measurable map $x:\Omega\to S$ is a random sequence in $\mathbb R^N$. Nevertheless we allow $(\Omega,\mathcal A)$ and $(R,\mathcal R)$ to be arbitrary measurable spaces in what follows, further specificity being superfluous. 

Given any measurable map $x=(x_t):\Omega\to S$, we introduce notation for three sequences of sigma-algebras on $\Omega$.
\begin{enumerate}[label=\upshape(\roman*)]
	\item For each $t\in\mathbb Z$ we denote by $\mathcal F_t(x)$ the sigma-algebra on $\Omega$ generated by the collection of maps $\{x_s:s\leq t\}$. We denote by $\mathcal F(x)$ the sequence of sigma-algebras $(\mathcal F_t(x))$.
	\item For each $t\in\mathbb Z$ we denote by $\mathcal B_t(x)$ the sigma-algebra on $\Omega$ generated by the collection of maps $\{x_s:s\geq t\}$.  We denote by $\mathcal B(x)$ the sequence of sigma-algebras $(\mathcal B_t(x))$.
	\item For each nonnegative $t\in\mathbb Z$ we denote by $\mathcal O_t(x)$ the sigma-algebra on $\Omega$ generated by the collection of maps $\{x_s:0\leq s\leq t\}$. For each negative $t\in\mathbb Z$ we denote by $\mathcal O_t(x)$ the sigma-algebra on $\Omega$ generated by the collection of maps $\{x_s:t\leq s\leq 0\}$. We denote by $\mathcal O(x)$ the sequence of sigma-algebras $(\mathcal O_t(x))$.
\end{enumerate}

We offer the following heuristic remarks on the interpretation of $\mathcal F(x)$, $\mathcal B(x)$ and $\mathcal O(x)$. In the statistical literature a sequence $(\mathcal A_t)$ of sub-sigma-algebras of $\mathcal A$ is called a filtration if $\mathcal A_t\subseteq\mathcal A_{t+1}$ for each $t\in\mathbb Z$. The sequence $\mathcal F(x)$ is a filtration in this sense. Adopting the conventional interpretation of a sigma-algebra as an information set, the requirement that $\mathcal A_t\subseteq\mathcal A_{t+1}$ for each $t\in\mathbb Z$ can be understood to mean that we recall the past. In this sense, time flows in a forward direction. In a world where time flows backward we instead recall the future, and the definition of a filtration is naturally modified to require that $\mathcal A_t\supseteq\mathcal A_{t+1}$ for each $t\in\mathbb Z$. The sequence $\mathcal B(x)$ is a filtration in this second sense. In a world where time flows outward from time zero we recall the events that transpire between time zero and the present moment, which could be termed the inward events, so the definition of a filtration is naturally modified to require that $\mathcal A_t\subseteq A_{t+1}$ for each nonnegative $t\in\mathbb Z$ and $\mathcal A_t\supseteq\mathcal A_{t+1}$ for each negative $t\in\mathbb Z$. The sequence $\mathcal O(x)$ is a filtration in this third sense. The three senses in which a sequence of sigma-algebras may be a filtration can be understood to correspond to three distinct arrows of time, these pointing forward, backward and outward.

A measurable map $x=(x_t):\Omega\to S$ is said to be \emph{adapted} to a sequence $(\mathcal A_t)$ of sub-sigma-algebras of $\mathcal A$ if $x_t$ is $\mathcal A_t$-measurable for each $t\in\mathbb Z$.

\begin{definition}\label{def:flow}
	Let $x:\Omega\to S$ and $y:\Omega\to S$ be measurable maps. We say that $y$ is
	\begin{enumerate}[label=\upshape(\roman*)]
		\item a \emph{forward} $x$\emph{-flow} if $y$ is adapted to $\mathcal F(x)$;
		\item a \emph{backward} $x$\emph{-flow} if $y$ is adapted to $\mathcal B(x)$;
		\item an \emph{outward} $x$\emph{-flow} if $y$ is adapted to $\mathcal O(x)$.
	\end{enumerate}
\end{definition}

Definition \ref{def:flow} is the basis for the labels given to the three sequences in \eqref{eq:forward}, \eqref{eq:backward} and \eqref{eq:outward}. When $\varepsilon$ is a random sequence in $\mathbb R^N$, these three sequences are, respectively, a forward $\varepsilon$-flow, a backward $\varepsilon$-flow, and an outward $\varepsilon$-flow. We understand each flow to be determined by its recollection of the sequence $\varepsilon$. The forward $\varepsilon$-flow recalls past values of $\varepsilon$, the backward $\varepsilon$-flow recalls future values of $\varepsilon$, and the outward $\varepsilon$-flow recalls inward values of $\varepsilon$.

In a world where the arrow of time points outward from time zero it is natural to understand time zero to be the origin of time. The concept of predetermination is easily understood in a world of this sort. If $x:\Omega\to S$ and $y:\Omega\to S$ are measurable maps and $y$ is an outward $x$-flow, then we may understand $y$ to be predetermined if $y$ is determined by the value taken by $x$ at time zero; that is, if $y$ is $\mathcal O_0(x)$-measurable. Put more prosaically, $y$ is predetermined if $y$ is determined by those of its recollections of $x$ which have been a part of its memory since the origin of time. In worlds where the arrow of time points forward or backward we might say that the origin of time is, respectively, minus or plus infinity. In such worlds we may again understand predetermination to mean that a sequence is determined by those of its recollections which have always been a part of its memory. The sigma-algebras
\begin{equation*}
	\mathcal F_{-\infty}(x)=\bigcap_{n=1}^\infty\mathcal F_{-n}(x)\quad\text{and}\quad\mathcal B_\infty(x)=\bigcap_{n=1}^\infty\mathcal B_n(x)
\end{equation*}
represent these recollections in a world where the arrow of time points respectively forward or backward.
\begin{definition}\label{def:predetflow}
	Let $x:\Omega\to S$ and $y:\Omega\to S$ be measurable maps. We say that $y$ is
	\begin{enumerate}[label=\upshape(\roman*)]
		\item a \emph{predetermined forward} $x$\emph{-flow} if $y$ is $\mathcal F_{-\infty}(x)$-measurable;
		\item a \emph{predetermined backward} $x$\emph{-flow} if $y$ is $\mathcal B_\infty(x)$-measurable;
		\item a \emph{predetermined outward} $x$\emph{-flow} if $y$ is $\mathcal O_0(x)$-measurable.
	\end{enumerate}
\end{definition}

By construction, a predetermined forward $x$-flow is necessarily a forward $x$-flow, and the same is true for the backward and outward directions. Definition \ref{def:predetflow} is the basis for the labels given to the three sequences in \eqref{eq:predetforward}, \eqref{eq:predetbackward} and \eqref{eq:predetoutward}. When $x$ is a random sequence in $\mathbb R^N$, these three sequences are, respectively, a predetermined forward $x$-flow, a predetermined backward $x$-flow, and a predetermined outward $x$-flow.

There is a long history in physics and philosophy of assigning an arrow to time based on considerations similar to those discussed in this section. The following passage from \citet{Eddington1929} has often been quoted.

\begin{quote}
	Let us draw an arrow arbitrarily. If as we follow the arrow we find more and more of the random element in the state of the world, then the arrow is pointing towards the future; if the random element decreases the arrow points towards the past. That is the only distinction known to physics.
\end{quote}

Our use of the terms forward, backward and outward in Definition \ref{def:flow} is consistent with Eddington's prescription. The association we have drawn between the arrow of time and the direction in which one's memory extends is also ground well-trodden. \citet{Hawking1988} defines the psychological arrow of time to be \emph{the direction of time in which we remember the past and not the future}, and compares this to Eddington's arrow of time, which he calls the thermodynamic arrow of time, arguing that the two are essentially the same. In our setting this is indeed the case, as the forward, backward and outward filtrations each simultaneously represent an accumulation of randomness and an accumulation of information. \citet{BarbourKoslowskiMercati2014} identify an outward arrow of time in a gravitational model, writing that \emph{it is very natural to identify an arrow of time with the direction in which structure (...) grows. We then have a dynamically enforced scenario with one past (...) and two futures}. One could say the same of an autoregressive law of motion with eigenvalues on the unit circle.

\begin{appendix}
		
	\section{Subexponentiality with probability one}\label{sec:prob1}
	
	Here we provide a sufficient condition for a random sequence in a normed space---for instance, the space $\mathbb R^N$---to be subexponential with probability one. Let $(\Omega,\mathcal A,\mu)$ be a probability space, let $\mathbb V$ be a normed space, and let $\mathbb S$ be the set of all sequences in $\mathbb V$ indexed by $t\in\mathbb Z$. We equip $\mathbb V$ with its Borel sigma-algebra and equip $\mathbb S$ with the corresponding product sigma-algebra, i.e.\ the coarsest sigma-algebra on $\mathbb S$ such that $s\mapsto s_t$ is a measurable map from $\mathbb S$ to $\mathbb V$ for each $t\in\mathbb Z$. 
	\begin{proposition}\label{prop:subexponential}
		Let $\varepsilon:\Omega\to\mathbb S$ be a measurable map. The set of all $\omega\in\Omega$ such that $\varepsilon(\omega)$ is subexponential belongs to $\mathcal A$. If
		\begin{align}
			\sum_{t\in\mathbb Z}r^{\lvert t\rvert}\int_\Omega\lVert \varepsilon_t\rVert\mathrm{d}\mu&<\infty\quad\text{for all }r\in(0,1)\label{eq:Esubexponential}
		\end{align}
		then $\mu\{\omega\in\Omega:\varepsilon(\omega)\text{ is subexponential}\}=1$.
	\end{proposition}
	\begin{proof}
		To see why $\{\omega\in\Omega:\varepsilon(\omega)\text{ is subexponential}\}\in\mathcal A$ we write
		\begin{multline*}
			\left\{\omega\in\Omega:\sum_{t\in\mathbb Z}r^{\lvert t\rvert}\lVert \varepsilon_t(\omega)\rVert<\infty\text{ for all }r\in(0,1)\right\}\\=\bigcap_{k=2}^\infty\bigcap_{\ell=1}^\infty\bigcup_{m=1}^\infty\bigcap_{n=1}^\infty\left\{\omega\in\Omega:\sum_{t=m}^{m+n}\left(1-\frac{1}{k}\right)^{t}\lVert \varepsilon_t(\omega)\rVert+\sum_{t=m}^{m+n}\left(1-\frac{1}{k}\right)^{t}\lVert \varepsilon_{-t}(\omega)\rVert<\frac{1}{\ell}\right\}
		\end{multline*}
		using Cauchy's criterion for the convergence of a series. The sets on the right-hand side all belong to $\mathcal A$ because $\lVert\varepsilon_t\rVert:\Omega\to\mathbb R$ is Borel measurable for each $t\in\mathbb Z$, a consequence of the measurability of $\varepsilon_t:\Omega\to\mathbb V$ and the continuity of $\lVert\cdot\rVert:\mathbb V\to\mathbb R$. We remain in $\mathcal A$ after taking countable unions and intersections.
		
		Condition \eqref{eq:Esubexponential} implies that
		\begin{align*}
			\mu\left\{\omega\in\Omega:\sum_{t\in\mathbb Z}r^{\lvert t\rvert}\lVert \varepsilon_t(\omega)\rVert<\infty\right\}&=1\quad\text{for all }r\in(0,1),
		\end{align*}
		because otherwise, by the monotone convergence theorem, we must have
		\begin{align*}
			\sum_{t\in\mathbb Z}r^{\lvert t\rvert}\int_\Omega\lVert\varepsilon_t\rVert\mathrm{d}\mu&=\int_\Omega\left(\sum_{t\in\mathbb Z}r^{\lvert t\rvert}\lVert\varepsilon_t\rVert\right)\mathrm{d}\mu=\infty\quad\text{for some }r\in(0,1),
		\end{align*}
		contradicting \eqref{eq:Esubexponential}. Therefore, if \eqref{eq:Esubexponential} is satisfied, then
		\begin{multline*}
			\mu\{\omega\in\Omega:\varepsilon(\omega)\text{ is subexponential}\}=\mu\bigcap_{n=2}^\infty\left\{\omega\in\Omega:\sum_{t\in\mathbb Z}\left(1-\frac{1}{n}\right)^{\lvert t\rvert}\lVert \varepsilon_t(\omega)\rVert<\infty\right\}\\
			=\lim_{n\to\infty}\mu\left\{\omega\in\Omega:\sum_{t\in\mathbb Z}\left(1-\frac{1}{n}\right)^{\lvert t\rvert}\lVert \varepsilon_t(\omega)\rVert<\infty\right\}=1,
		\end{multline*}
		using the continuity from above property of probabilities.
	\end{proof}
		
\end{appendix}

\end{document}